\documentclass{aa}
\usepackage{psfig}

\begin{document}

\title{Unveiling the nature of {\it INTEGRAL} objects through optical 
spectroscopy. V. Identification and properties of 21 southern hard X--ray 
sources\thanks{Based on observations collected at the Cerro Tololo 
Interamerican Observatory (Chile).}}

\author{N. Masetti\inst{1},
L. Morelli\inst{2},
E. Palazzi\inst{1},
G. Galaz\inst{2},
L. Bassani\inst{1},
A. Bazzano\inst{3}, 
A.J. Bird\inst{4},
A.J. Dean\inst{4},
G.L. Israel\inst{5},
R. Landi\inst{1},
A. Malizia\inst{1},
D. Minniti\inst{2},
F. Schiavone\inst{1},
J.B. Stephen\inst{1},
P. Ubertini\inst{3} and
R. Walter\inst{6}
}

\institute{
INAF -- Istituto di Astrofisica Spaziale e Fisica Cosmica di 
Bologna, Via Gobetti 101, I-40129 Bologna, Italy (formerly IASF/CNR,
Bologna)
\and
Departamento de Astronom\'{i}a y Astrof\'{i}sica, Pontificia Universidad 
Cat\'olica de Chile, Casilla 306, Santiago 22, Chile
\and
INAF -- Istituto di Astrofisica Spaziale e Fisica Cosmica di
Roma, Via Fosso del Cavaliere 100, I-00133 Rome, Italy (formerly 
IASF/CNR, Rome)
\and
School of Physics \& Astronomy, University of Southampton, Southampton, 
Hampshire, SO17 1BJ, United Kingdom  
\and
INAF - Osservatorio Astronomico di Roma, via Frascati 33, I-00040 
Monteporzio Catone, Italy
\and
INTEGRAL Science Data Centre, Chemin d'Ecogia 16, CH-1290 Versoix,
Switzerland
}

\offprints{N. Masetti (\texttt{masetti@iasfbo.inaf.it)}}
\date{Received 17 July 2006; accepted 2 August 2006}

\abstract{
Optical spectroscopic identification of the nature of 21 unidentified 
southern hard X--ray objects is reported here in the framework of our 
campaign aimed at determining the nature of newly-discovered and/or 
unidentified sources detected by {\it INTEGRAL}.
Our results show that 5 of these objects are magnetic Cataclysmic 
Variables (CVs), 4 are High-Mass X--ray Binaries (HMXBs; one of which 
is in the Large Magellanic Cloud) and 12 are Active Galactic Nuclei 
(AGNs).
When feasible, the main physical parameters for these hard X--ray sources 
are also computed using the multiwavelength information available in the 
literature. These identifications further underscore the importance of 
{\it INTEGRAL} in the study of the hard X--ray spectrum of AGNs, HMXBs and 
CVs, and the usefulness of a strategy of catalogues cross-correlation plus 
optical spectroscopy to securely pinpoint the actual nature of the X--ray 
sources detected with {\it INTEGRAL}.
}

\titlerunning{The nature of 21 hard X--ray {\it INTEGRAL} sources}
\authorrunning{N. Masetti et al.}

\maketitle

\section{Introduction}

\begin{figure*}%[t!]
\vspace{-2cm}
\hspace{-1.7cm}
%\centering{\mbox{
\psfig{file=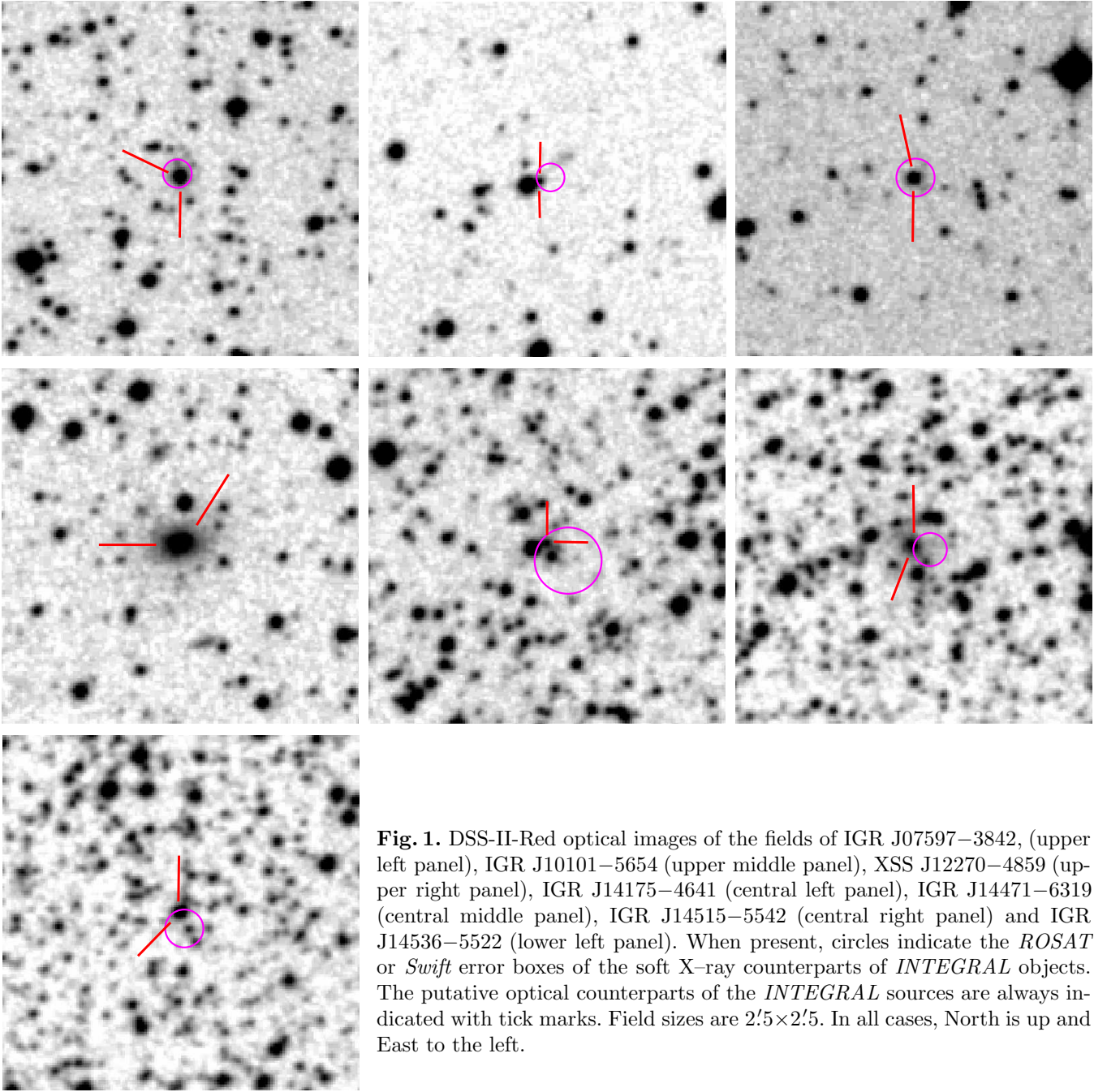,width=21cm}%}}

%\begin{center}
%\hspace{-.1cm}
%\centering{\mbox{\psfig{file=igr07597.ps,width=5.9cm}}}
%\vspace{-.3cm}
%\centering{\mbox{\psfig{file=igr10101.ps,width=5.9cm}}}
%\vspace{-.3cm}
%\centering{\mbox{\psfig{file=xss12270.ps,width=5.9cm}}}
%\vspace{-.3cm}
%\centering{\mbox{\psfig{file=igr14175.ps,width=5.9cm}}}
%\vspace{-.3cm}
%\centering{\mbox{\psfig{file=igr14471.ps,width=5.9cm}}}
%\vspace{-.3cm}
%\centering{\mbox{\psfig{file=igr14515.ps,width=5.9cm}}}
%\vspace{-.3cm}
%\begin{center}
\vspace{-13.7cm}
\parbox{6.0cm}{
\hspace{6.0cm}
%\psfig{file=igr14536.ps,width=5.9cm}
%\end{center}
%\vspace{-1cm}
}
%\hspace{0.2cm}
\parbox{11.7cm}{
%\vspace{-8.5cm}
\caption{DSS-II-Red optical images of the fields of IGR J07597$-$3842,
(upper left panel), IGR J10101$-$5654 (upper middle panel), XSS 
J12270$-$4859 (upper right panel), IGR J14175$-$4641 (central left 
panel), IGR J14471$-$6319 (central middle panel), IGR J14515$-$5542
(central right panel) and IGR J14536$-$5522 (lower left panel). 
When present, circles indicate the {\it ROSAT} or {\it Swift} error 
boxes of the soft X--ray counterparts of {\it INTEGRAL} objects.
The putative optical counterparts of the {\it INTEGRAL} sources 
are always indicated with tick marks. Field sizes are 
2$\farcm$5$\times$2$\farcm$5.
In all cases, North is up and East to the left.}
\vspace{1cm}
}
%\end{center}
\end{figure*}

\begin{figure*}%[th!]
\vspace{-2cm}
\hspace{-1.7cm}
%\centering{\mbox{
\psfig{file=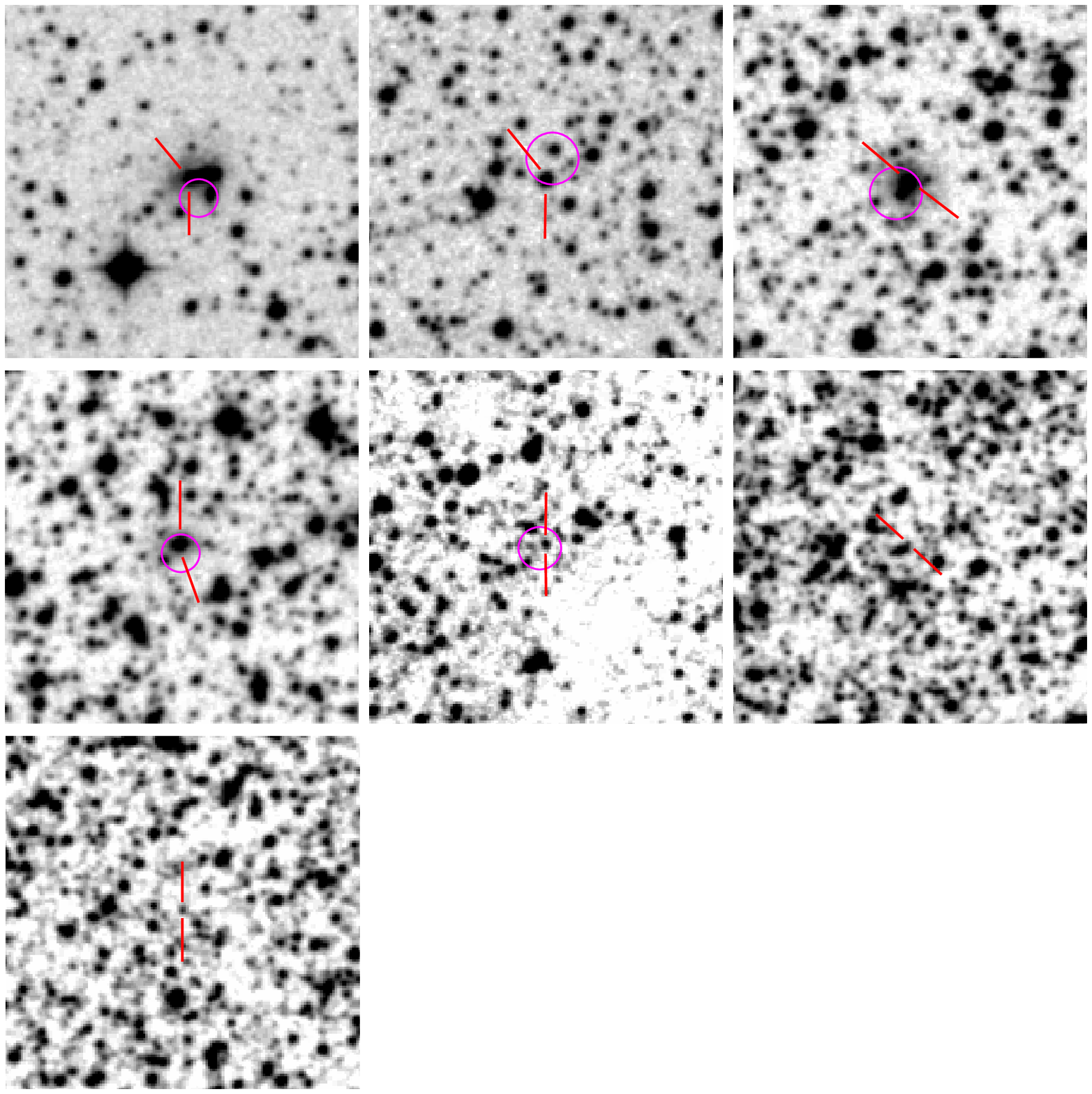,width=21cm}%}}

%\begin{center}
%\hspace{-.1cm}
%\centering{\mbox{\psfig{file=igr14552.ps,width=5.9cm}}}
%\vspace{-.3cm}
%\centering{\mbox{\psfig{file=igr15094.ps,width=5.9cm}}}
%\vspace{-.3cm}
%\centering{\mbox{\psfig{file=igr16185.ps,width=5.9cm}}}
%\vspace{-.3cm}
%\centering{\mbox{\psfig{file=igr16558.ps,width=5.9cm}}}
%\vspace{-.3cm}
%\centering{\mbox{\psfig{file=igr17200.ps,width=5.9cm}}}
%\vspace{-.3cm}
%\centering{\mbox{\psfig{file=igr17488.ps,width=5.9cm}}}
%\vspace{-.3cm}
%\begin{center}
%\vspace{-0.1cm}
\vspace{-13.5cm}
\parbox{6.0cm}{
\hspace{6cm}
%\psfig{file=igr17513.ps,width=5.9cm}
%\end{center}
%\vspace{-1cm}
}
%\hspace{0.2cm}
\parbox{11.7cm}{
%\vspace{-8.5cm}
\caption{As Fig. 1, but for the fields of IGR J14552$-$5133,
(upper left panel), IGR J15094$-$6649 (upper middle panel), IGR  
J16185$-$5928 (upper right panel), IGR J16558$-$5203 (central left 
panel), IGR J17200$-$3116 (central middle panel), IGR J17488$-$3253
(central right panel) and IGR J17513$-$2011 (lower left panel). 
For the sake of clarity, the {\it Chandra} error circles of the
latter two sources were not reported in the corresponding panels.}
\vspace{1.5cm}
}
%\end{center}
\end{figure*}

Since its launch in October 2002, the {\it INTEGRAL} satellite (Winkler et 
al. 2003) is boosting our knowledge of the hard X--ray sky above 20 
keV in terms of both sensitivity and positional accuracy of the detected 
sources. Thanks to the capabilities of the IBIS instrument (Ubertini
et al. 2003), {\it INTEGRAL} is effectively 
detecting hard X--ray objects at the mCrab level with a typical 
localization accuracy of 2--3$'$ (Gros et al. 2003). This has made it 
possible, for the first time, to obtain all-sky maps in the 20--100 keV 
range with arcminute accuracy and down to mCrab sensitivities (e.g., Bird 
et al. 2006).

Most of the sources detected by {\it INTEGRAL} are known Galactic X--ray 
binaries ($\sim$50\% of the total number of detected objects), plus a 
fraction of known Active Galactic Nuclei (AGNs; $\sim$10\%) and 
Cataclysmic Variables (CVs; $\sim$5\%). However, a large majority of 
the remaining objects (about 25\% of all detections achieved with IBIS) 
has no obvious counterpart at other wavelengths and therefore cannot 
immediately be associated with any known class of high-energy emitting 
objects.

Recently, in order to fill this identification gap, we started a campaign 
aimed at identifying the nature of these still unknown sources through 
optical spectroscopy at northern and southern telescopes (Masetti et al. 
2004, 2006a,b,c; hereafter Papers I-IV). Our results indicate that, 
despite {\it INTEGRAL} doubled the number of Galactic High Mass X--ray 
Binaries (HMXBs; see Walter et al. 2006) and despite the expectation 
according to which most of these unidentified objects should be 
HMXBs (Dean et al. 2005), about half of them are actually 
identified in the optical as nearby ($z \la$ 0.1) AGNs (Papers I-IV).

In the framework of our continuing effort to identify unknown {\it 
INTEGRAL} sources, we present here the optical spectroscopic observations 
obtained on 21 southern objects at the 1.5-metre telescope of the Cerro 
Tololo Interamerican Observatory (CTIO) located in Cerro Tololo (Chile). 
In Sect. 2 we introduce the criteria with which the sample of {\it 
INTEGRAL} and optical objects was chosen for the present observational 
campaign, whereas in Sect. 3 a description of the observations is given; 
Sect. 4 reports and discusses the results, divided into three broad 
classes of sources (CVs, HMXBs and AGNs), together with a statistical 
outline of the identifications of {\it INTEGRAL} sources obtained up to 
now. Conclusions are drawn in Sect. 5.

\section{Sample selection}

In order to continue our program (Papers I-IV) of optical spectroscopic 
identifications of {\it INTEGRAL} sources with unknown nature, we first 
collected all objects belonging to the 2$^{\rm nd}$ IBIS Galactic 
Plane Survey (Bird et al. 2006), to the Crux arm Tangent Survey (Revnivtsev 
et al. 2006a), to the AGN minisurvey of Sazonov et al. (2005) and to the 
Circinus-Carina arm Survey (Kuiper et al. 2006), and which are visible 
from the southern hemisphere.

We then positionally cross-correlated the IBIS error circles of the 
selected southern unidentified {\it INTEGRAL} objects with catalogues 
of soft ($<$10 keV) 
X--ray sources. This was made in order to reduce the X--ray error box size 
to some ($\la$10) arcsec at most. For the present sample, we selected {\it 
INTEGRAL} objects with a single {\it ROSAT} source (Voges et al. 1999, 
2000; {\it ROSAT} Team 2000), or a single {\it Swift}/XRT archival X--ray 
source (available at {\tt http://www.asdc.asi.it}), or a single {\it 
Chandra} source (Sazonov et al. 2005; Halpern 2005; Tomsick et al. 2006; 
Israel et al., in preparation) within the IBIS error box. This approach 
was chosen because Stephen et al. (2005, 2006) show that, from a 
statistical argument, these are very likely to be the soft X--ray 
counterparts of the positionally corresponding {\it INTEGRAL} sources; 
besides, the results of Papers I-IV prove that this approach is very 
effective, when combined with optical spectroscopy.

For the cross-correlation searches, we considered 90\% confidence level 
{\it INTEGRAL}/IBIS error circles. To this aim, a conservative 90\% 
confidence level error box radius of 2$'$ was assumed for the objects 
belonging to the 2$^{\rm nd}$ IBIS Galactic Plane Survey (Bird et al. 
2006), and of 6$'$ for the Crux arm Survey objects (as stated in Revnivtsev 
et al. 2006a). For the Circinus-Carina Survey sources, the 90\% confidence 
level error box radius as reported in Kuiper et al. (2006) regarding each 
object was considered.

In this way we could select 18 unidentified {\it INTEGRAL} sources 
associated with a single arcsec-sized soft X--ray error box which,
when overlaid onto the corresponding DSS-II-Red 
survey\footnote{available at {\tt http://archive.eso.org/dss/dss}}
images, is seen to contain a single or few (3 at most) relatively bright 
($R <$ 18) possible optical counterparts.
Three additional sources (IGR J14175$-$4641, IGR J14552$-$5133 and IGR 
J18244$-$5622) were added to our sample as their IBIS error circle
includes bright field objects which were suggested as their possible 
counterparts (Revnivtsev et al. 2006a,b).
We refer the reader to Paper III for the caveats of choosing, within the 
IBIS error box, ``peculiar" sources which are not readily associated 
with an arcsec-sized soft X--ray position.

The list of selected {\it INTEGRAL} sources is shown in the leftmost
column of Table 1. Figures 1 and 2 report the optical DSS-II-Red 
survey fields of the 14 sources (out of the 21 of the selected sample) 
for which no optical finding chart was published before the present work. 
In these figures, the position of the proposed counterpart of each 
{\it INTEGRAL} source is also shown (see also Table 2).
The finding charts of the remaining sources can be found in Sazonov et al. 
(2005), Tomsick et al. (2006) and Revnivtsev et al. (2006b).

In the following, when not explicitly stated otherwise, for our X--ray 
flux estimates we will assume a Crab-like spectrum. We also remark that 
the results presented here supersede the preliminary ones of Masetti et 
al. (2006d).

\section{Optical spectroscopy}

All objects were observed spectroscopically with the 1.5-metre CTIO 
telescope of Cerro Tololo (Chile) equipped with the R-C spectrograph, 
which carries a 1274$\times$280 pixels Loral CCD. Data were acquired using 
Grating \#13/I and with a slit width of 1$\farcs$5, giving a nominal 
spectral coverage between 3300 and 10500 \AA~and a dispersion of 
5.7~\AA/pix. The complete log of the observations is reported in Table 1.

After cosmic-ray rejection, the spectra were reduced, background 
subtracted and optimally extracted (Horne 1986) using IRAF\footnote{IRAF 
is the Image Reduction and Analysis Facility made available to the 
astronomical community by the National Optical Astronomy Observatories, 
which are operated by AURA, Inc., under contract with the U.S. National 
Science Foundation. It is available at {\tt http://iraf.noao.edu/}}. 
Wavelength calibration was performed using He-Ar lamps acquired soon after 
each spectroscopic exposure; the spectra were then flux-calibrated using 
the spectrophotometric standards LTT 3218 and LTT 7379 (Hamuy et al. 1992, 
1994).

Finally, and when applicable, different spectra of the same object were
stacked together to increase the S/N ratio. The wavelength calibration
uncertainty was $\sim$0.5~\AA~for all cases; this was checked using
the positions of background night sky lines.

\begin{table*}[th!]
\caption[]{Log of the spectroscopic observations presented in this paper.}
\begin{center}
\begin{tabular}{llcc}
\noalign{\smallskip}
\hline
\hline
\noalign{\smallskip}
\multicolumn{1}{c}{Object} & \multicolumn{1}{c}{Date} & Mid-exposure & 
Exposure \\
 & & time (UT) & time (s) \\
\noalign{\smallskip}
\hline
\noalign{\smallskip}

IGR J05007$-$7047    & 21 Mar 2006 & 00:35:17 & 2$\times$1200 \\
IGR J07565$-$4139    & 21 Mar 2006 & 01:42:08 & 2$\times$1200 \\
IGR J07597$-$3842    & 21 Mar 2006 & 02:39:12 & 2$\times$1200 \\
IGR J10101$-$5654    & 06 Apr 2006 & 03:05:43 & 2$\times$1800 \\
IGR J12026$-$5349    & 21 Mar 2006 & 03:46:53 & 2$\times$1800 \\
XSS J12270$-$4859    & 22 Mar 2006 & 01:32:07 & 2$\times$1800 \\
IGR J14175$-$4641    & 21 Mar 2006 & 04:44:08 & 2$\times$1200 \\
IGR J14471$-$6319    & 06 Apr 2006 & 04:23:39 & 2$\times$1800 \\
IGR J14515$-$5542    & 22 Mar 2006 & 03:21:20 & 2$\times$1800 \\
IGR J14536$-$5522    & 22 Mar 2006 & 04:21:53 & 2$\times$1200 \\
IGR J14552$-$5133    & 22 Mar 2006 & 05:40:02 &          1800 \\
IGR J15094$-$6649    & 22 Mar 2006 & 06:50:21 &     1800+1200 \\
IGR J16167$-$4957    & 21 Mar 2006 & 06:03:46 & 2$\times$1800 \\
IGR J16185$-$5928    & 22 Mar 2006 & 08:02:00 & 2$\times$1800 \\
IGR J16207$-$5129    & 23 Mar 2006 & 06:33:05 & 2$\times$1800 \\
IGR J16558$-$5203    & 21 Mar 2006 & 07:26:17 & 2$\times$1200 \\
IGR J17195$-$4100    & 21 Mar 2006 & 08:39:00 & 2$\times$1800 \\
IGR J17200$-$3116    & 06 Apr 2006 & 05:34:33 & 2$\times$1500 \\
IGR J17488$-$3253    & 04 Apr 2006 & 07:11:21 & 2$\times$2400 \\
IGR J17513$-$2011    & 04 Apr 2006 & 08:53:27 & 2$\times$2700 \\
IGR J18244$-$5622    & 05 Apr 2006 & 09:35:21 & 2$\times$1200 \\

\noalign{\smallskip}
\hline
\hline
\noalign{\smallskip}
\end{tabular}
\end{center}
\end{table*}

\section{Results}

In this section we present the results of our spectroscopic campaign at 
CTIO. The optical magnitudes quoted below, if not otherwise stated, are 
extracted from the USNO-A2.0\footnote{available at \\ {\tt 
http://archive.eso.org/skycat/servers/usnoa/}} catalogue.

For the determination of the distance of compact Galactic X--ray sources,
in the case of CVs we will assume an absolute magnitude M$_V \sim$ 9 and 
an intrinsic color index $(V-R)_0 \sim$ 0 mag (Warner 1995), whereas for 
HMXBs, when applicable, we will use the intrinsic stellar color indices 
and absolute magnitudes as reported in Lang (1992) and Wegner (1994).

When applicable,
for the calculation of the absorption local to an AGN host galaxy,
we first dereddened the H$_\alpha$ and H$_\beta$ line 
fluxes by applying a correction for the Galactic absorption along the 
source line of sight. This was done following the prescription for the 
computation of the Galactic color excess $E(B-V)_{\rm Gal}$ as in 
Schlegel et al. (1998), and considering the Galactic extinction law of 
Cardelli et al. (1989).
Then, we assumed an intrinsic H$_\alpha$/H$_\beta$ line ratio of 
2.86 (Osterbrock 1989) and we computed the color excess 
$E(B-V)_{\rm AGN}$ local to the AGN host, using again the Cardelli et 
al.'s (1989) extinction law, from the comparison between the intrinsic 
line ratio and the one corrected for the Galactic reddening.

The spectra of the galaxies shown here were not corrected for
starlight contamination (see, e.g., Ho et al. 1993, 1997) given their
limited S/N and resolution. We do not consider this to affect any of our
conclusions. In the following we assume a cosmology with $H_{\rm 0}$ = 65
km s$^{-1}$ Mpc$^{-1}$, $\Omega_{\Lambda}$ = 0.7 and $\Omega_{\rm m}$ =
0.3. 

For the AGN classification, we used the criteria of Veilleux \& Osterbrock 
(1987) and the line ratio diagnostics of Ho et al. (1993, 1997); moreover,
for the subclass assignation of Seyfert 1 nuclei, we used the 
H$_\beta$/[O{\sc iii}]$\lambda$5007 line flux ratio criterion
as per Winkler (1992).

In Table 2 we report the equatorial J2000 coordinates of the proposed 
optical counterparts lacking up to now a precise (i.e. at one-arcsecond 
accuracy or better) position; we also show these objects in Figs. 1 and 2. 
This information is extracted from the 2MASS catalogue (with $\la$0$\farcs$1 
accuracy, according to Skrutskie et al. 2006) but for IGR J14552$-$5133, for 
which the DSS-II-Red astrometry (which has $\sim$1$''$ accuracy) was used
as no entry is present for this source in the 2MASS catalogue. The 
accurate coordinates of the 7 objects not included in Table 2 are 
reported in Tomsick et al. (2006), Sazonov et al. (2005) and 
Revnivtsev et al. (2006b).

In the following Subsections we give the object identifications divided 
into three broad classes (CVs, HMXBs and AGNs) ordered according to their 
increasing distance from Earth.

\begin{table}%[h!]
\caption[]{Equatorial J2000 coordinates of the counterparts to the
{\it INTEGRAL} sources of the sample presented in this paper and lacking 
an accurate (arcsecond or better) position up to now. Coordinates (from 
the 2MASS catalogue) have an accuracy better than 0$\farcs$1, with the
exception of those for IGR J14552$-$5133, which were extracted from 
the DSS-II-Red astrometry and have $\sim$1$''$ accuracy.}
\begin{center}
\begin{tabular}{lll}
\noalign{\smallskip}
\hline
\hline
\noalign{\smallskip}
\multicolumn{1}{c}{Object} & \multicolumn{1}{c}{RA (J2000)} & 
\multicolumn{1}{c}{Dec (J2000)} \\
\noalign{\smallskip}
\hline
\noalign{\smallskip}

IGR J07597$-$3842 & 07$^{\rm h}$ 59$^{\rm m}$ 41$\fs$819 & 
$-$38$^\circ$ 43$'$ 56$\farcs$03 \\
IGR J10101$-$5654 & 10$^{\rm h}$ 10$^{\rm m}$ 11$\fs$866 & 
$-$56$^\circ$ 55$'$ 32$\farcs$06  \\
XSS J12270$-$4859 & 12$^{\rm h}$ 27$^{\rm m}$ 58$\fs$748 & 
$-$48$^\circ$ 53$'$ 42$\farcs$88 \\
IGR J14175$-$4641 & 14$^{\rm h}$ 17$^{\rm m}$ 03$\fs$662 & 
$-$46$^\circ$ 41$'$ 41$\farcs$19 \\
IGR J14471$-$6319 & 14$^{\rm h}$ 47$^{\rm m}$ 14$\fs$881 & 
$-$63$^\circ$ 17$'$ 19$\farcs$24 \\
IGR J14515$-$5542 & 14$^{\rm h}$ 51$^{\rm m}$ 33$\fs$131 & 
$-$55$^\circ$ 40$'$ 38$\farcs$40 \\
IGR J14536$-$5522 & 14$^{\rm h}$ 53$^{\rm m}$ 41$\fs$055 & 
$-$55$^\circ$ 21$'$ 38$\farcs$74 \\
IGR J14552$-$5133 & 14$^{\rm h}$ 55$^{\rm m}$ 17$\fs$8 & 
$-$51$^\circ$ 34$'$ 17$''$ \\
IGR J15094$-$6649 & 15$^{\rm h}$ 09$^{\rm m}$ 26$\fs$013 & 
$-$66$^\circ$ 49$'$ 23$\farcs$29 \\
IGR J16185$-$5928 & 16$^{\rm h}$ 18$^{\rm m}$ 36$\fs$441 & 
$-$59$^\circ$ 27$'$ 17$\farcs$36 \\
IGR J16558$-$5203 & 16$^{\rm h}$ 56$^{\rm m}$ 05$\fs$618 & 
$-$52$^\circ$ 03$'$ 40$\farcs$87 \\
IGR J17200$-$3116 & 17$^{\rm h}$ 20$^{\rm m}$ 05$\fs$913 & 
$-$31$^\circ$ 16$'$ 59$\farcs$65 \\
IGR J17488$-$3253 & 17$^{\rm h}$ 48$^{\rm m}$ 55$\fs$129 & 
$-$32$^\circ$ 54$'$ 52$\farcs$15 \\
IGR J17513$-$2011 & 17$^{\rm h}$ 51$^{\rm m}$ 13$\fs$623 & 
$-$20$^\circ$ 12$'$ 14$\farcs$58 \\

\noalign{\smallskip}
\hline
\hline
\noalign{\smallskip}
\end{tabular}
\end{center}
\end{table}

\subsection{CVs}

\begin{figure*}[th!]
%\begin{center}
%\hspace{.1cm}
\hspace{-.8cm}
%\centering{
\mbox{\psfig{file=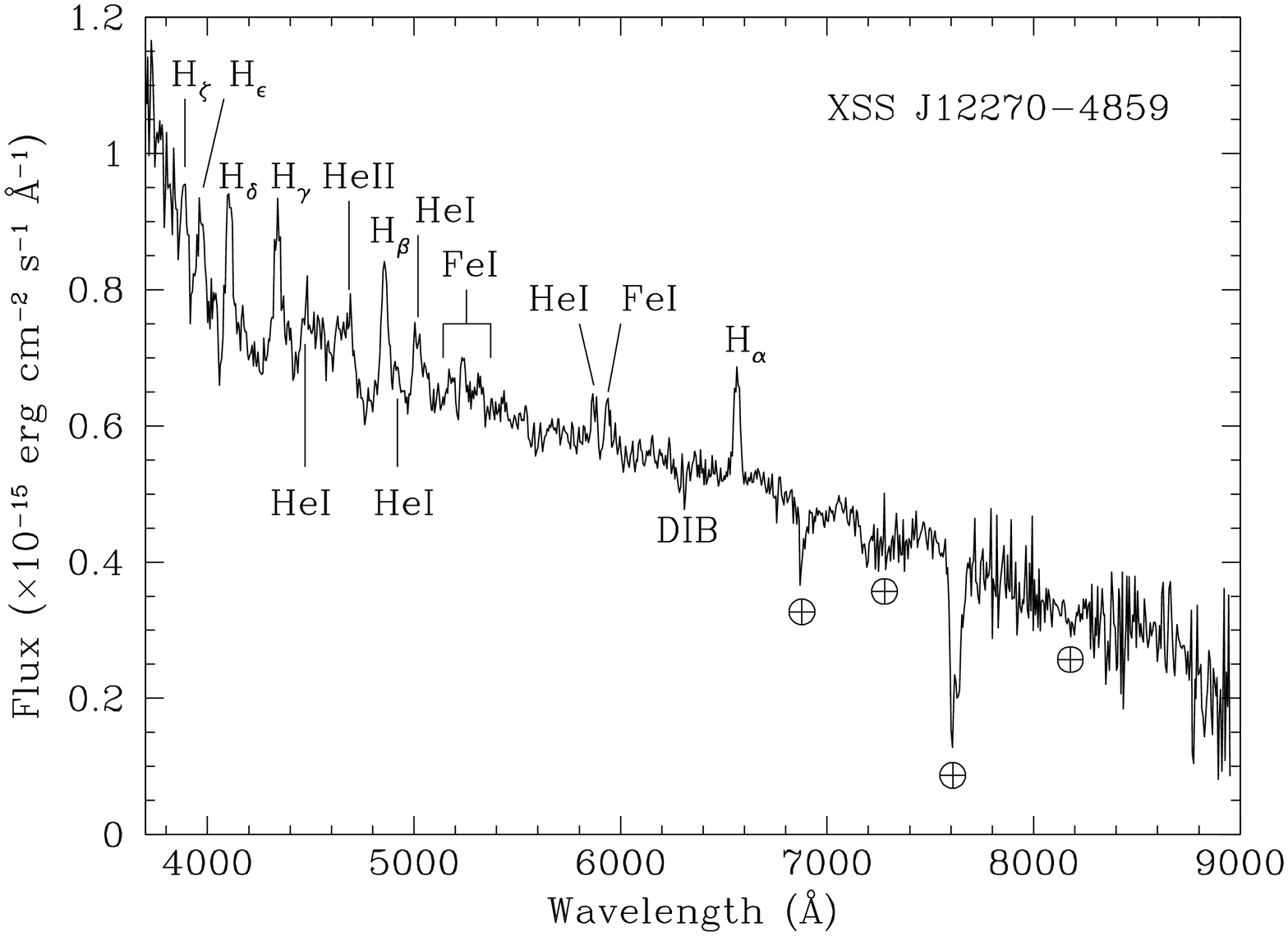,width=10.1cm,angle=270}}%}
\hspace{-1.0cm}
%\centering{
\mbox{\psfig{file=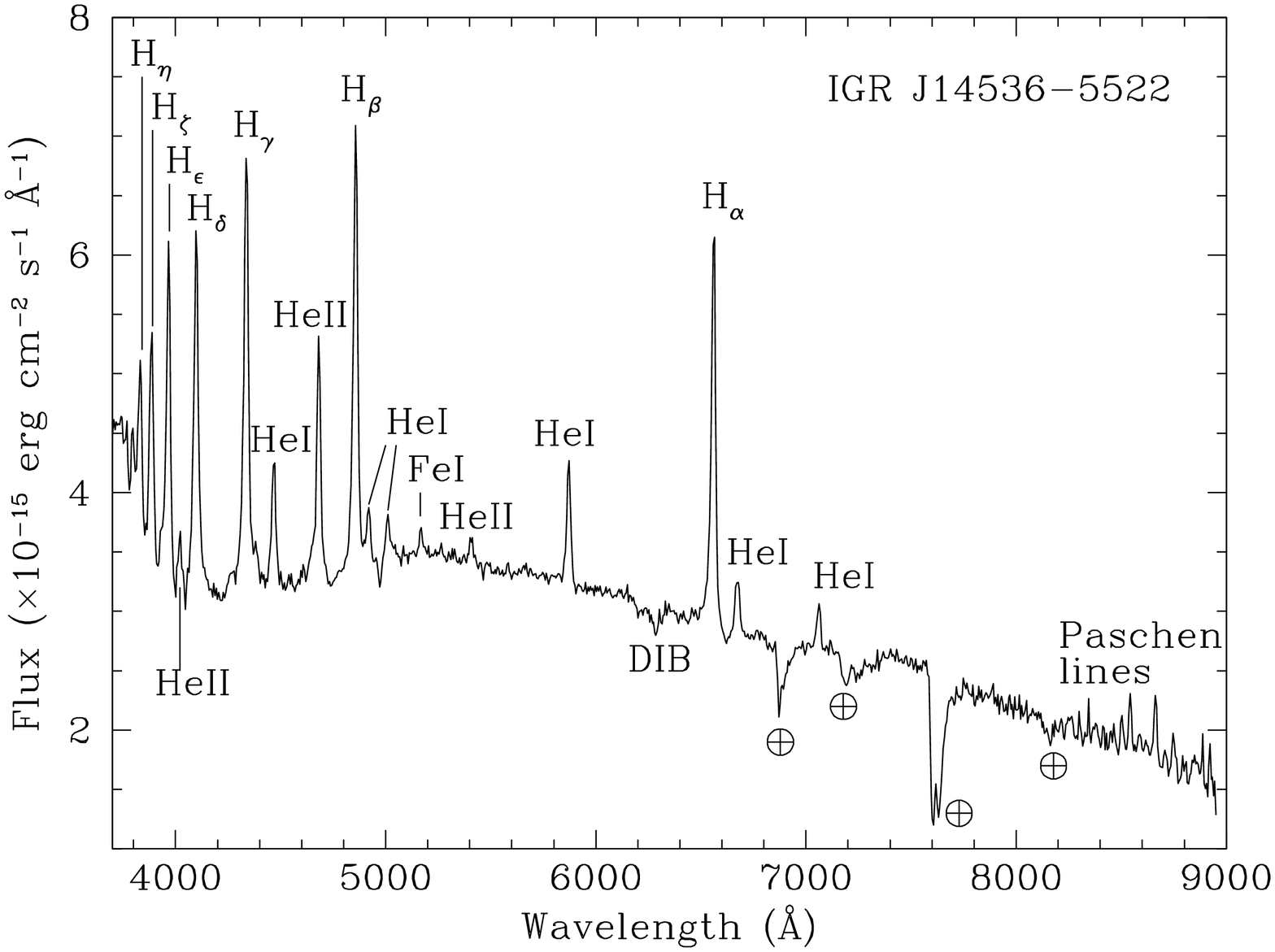,width=10.1cm,angle=270}}%}

\hspace{-.8cm}
%\centering{
\mbox{\psfig{file=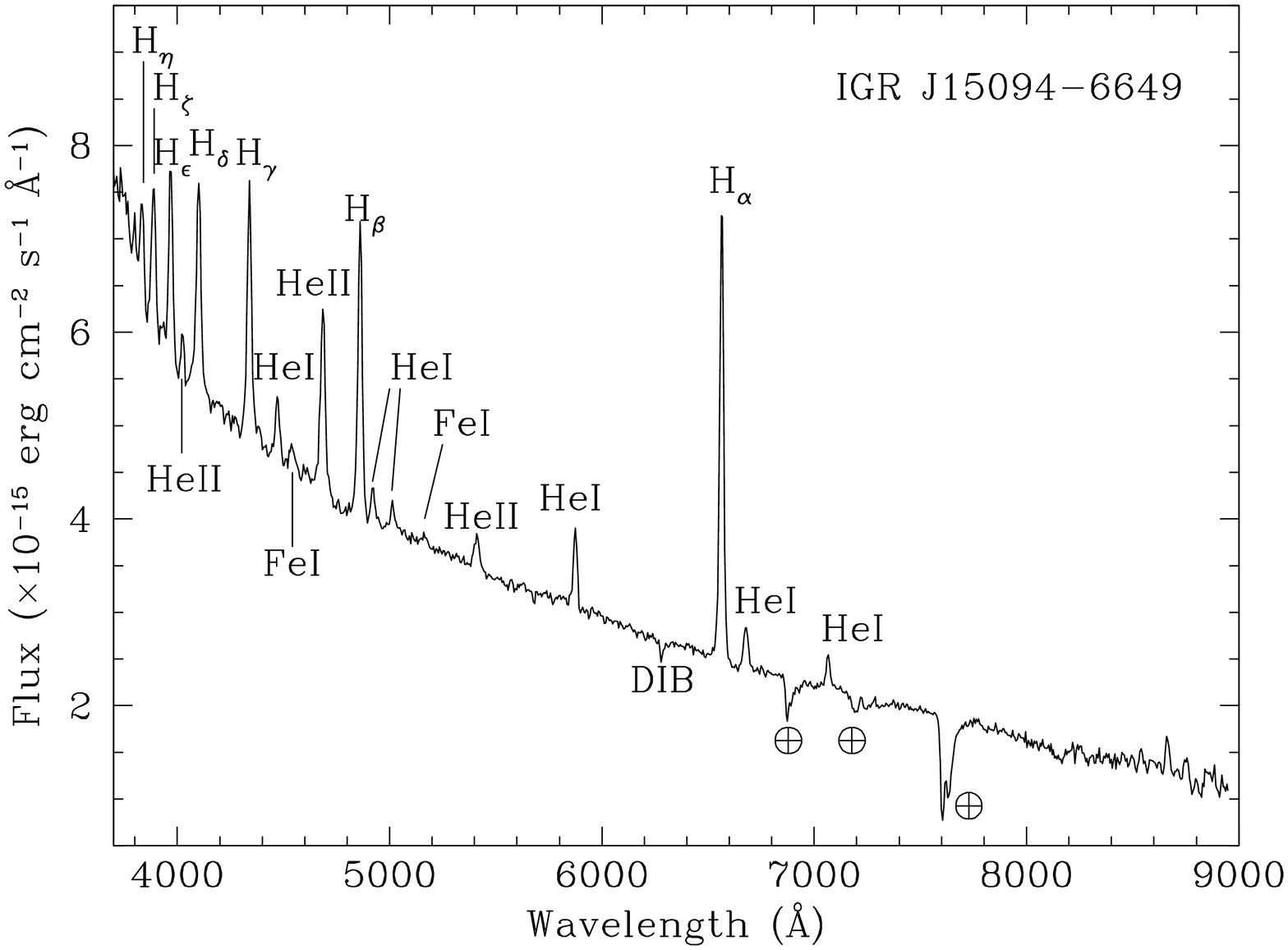,width=10.1cm,angle=270}}%}
\hspace{-1cm}
%\centering{
\mbox{\psfig{file=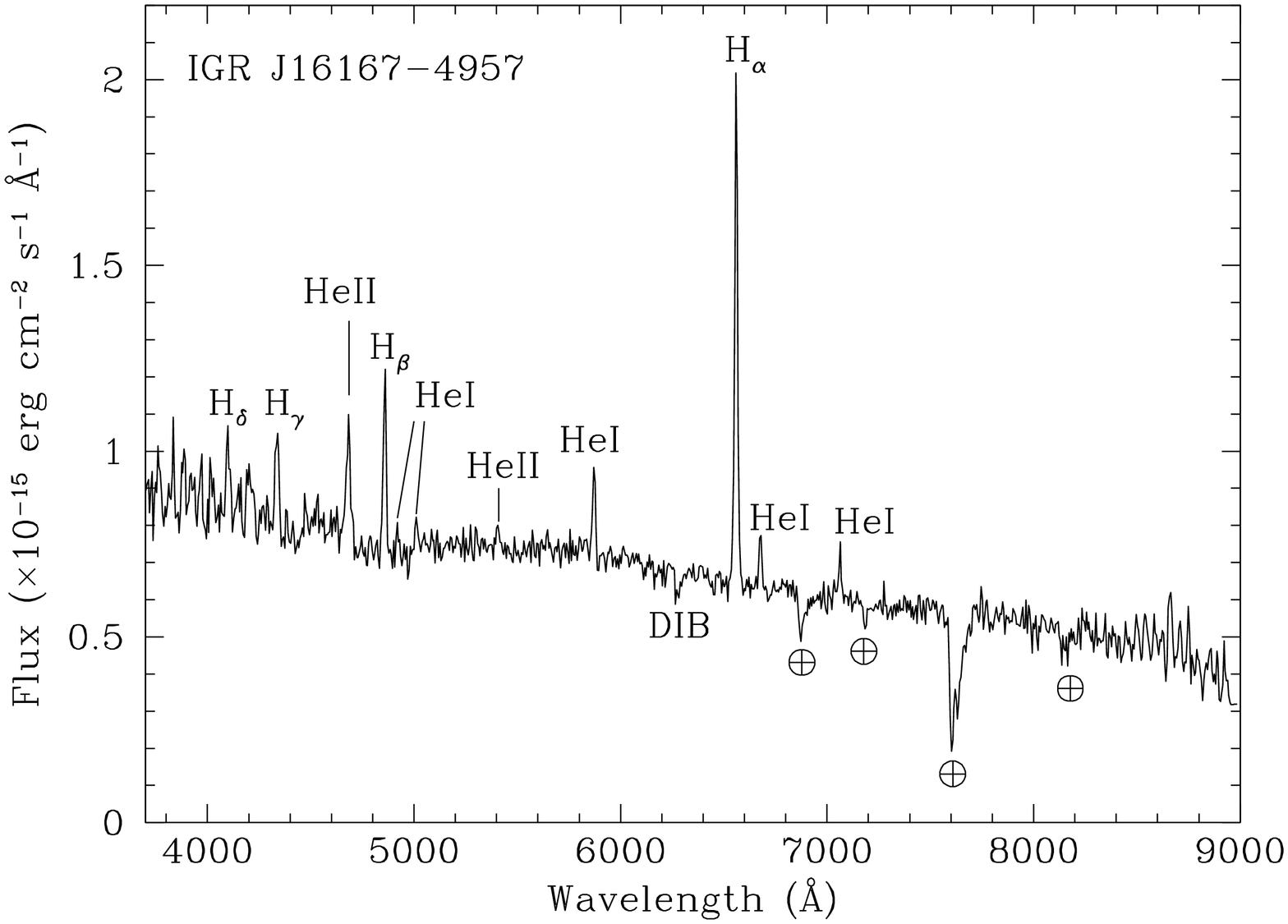,width=10.1cm,angle=270}}%}

%\begin{center}
\hspace{-.8cm}
\parbox{9cm}{
\psfig{file=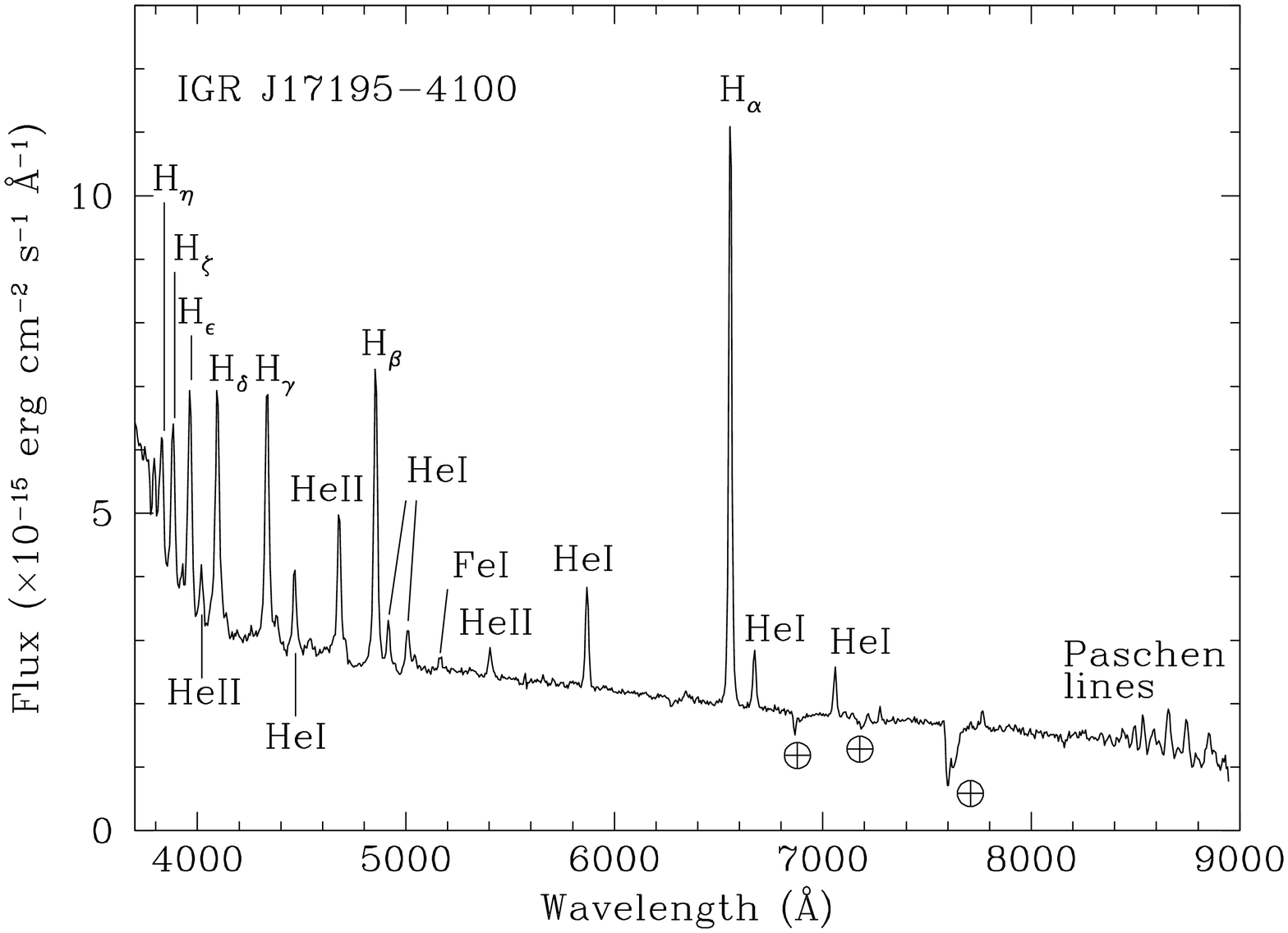,width=10.1cm,angle=270}
}
\hspace{0.5cm}
\parbox{7.5cm}{
\vspace{-1cm}
\caption{Spectra (not corrected for the intervening Galactic absorption) 
of the optical counterparts of the CVs identified in this sample;
all spectra were acquired with the 1.5m CTIO 
telescope. For each spectrum the main spectral features are labeled. The 
symbol $\oplus$ indicates atmospheric telluric absorption bands.}
}
%\end{center}
\end{figure*}

\begin{figure*}[th!]
%\begin{center}
%\hspace{.1cm}
\hspace{-.8cm}
%\centering{
\mbox{\psfig{file=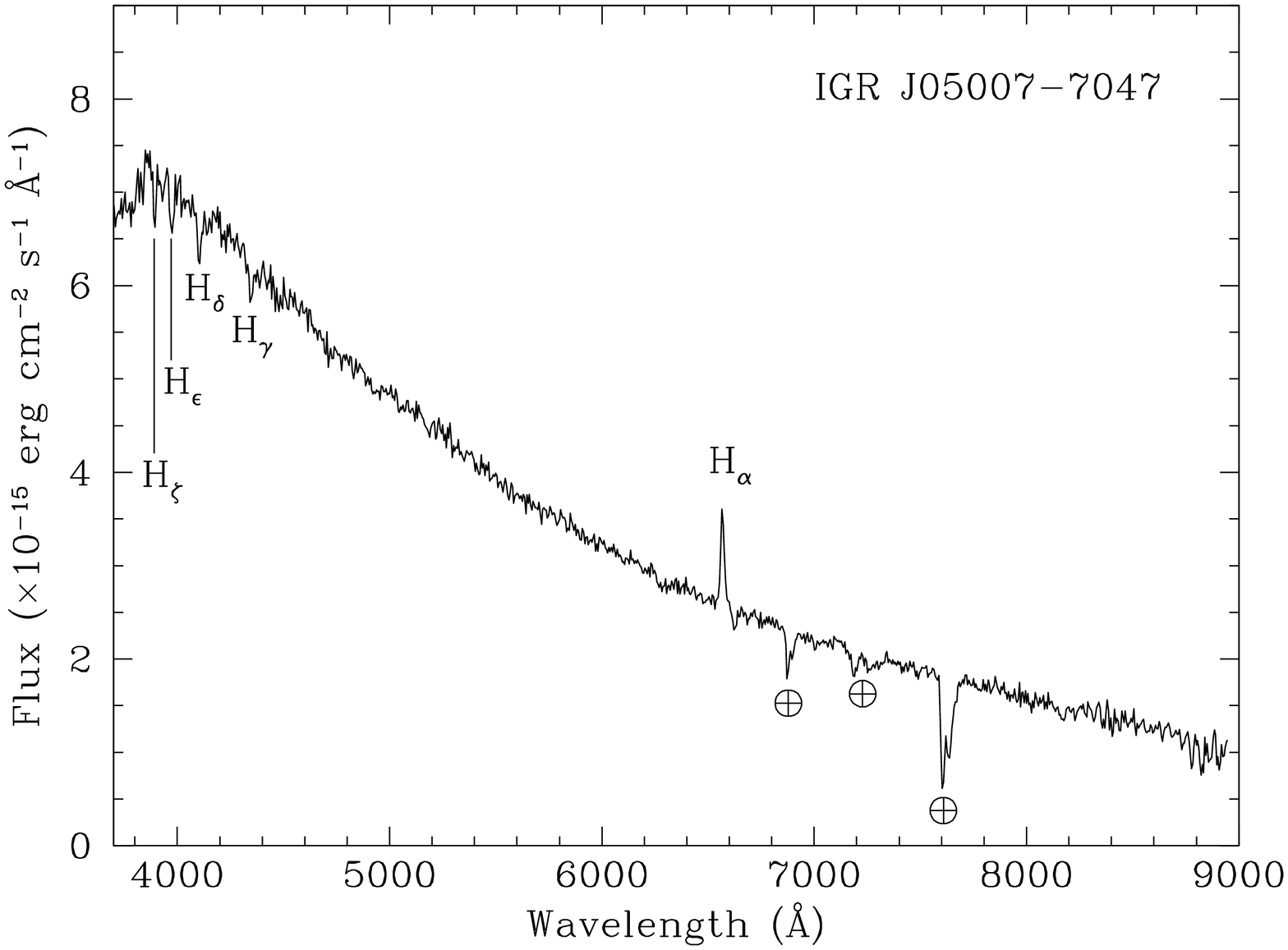,width=10.1cm,angle=270}}%}
\hspace{-1.0cm}
%\centering{
\mbox{\psfig{file=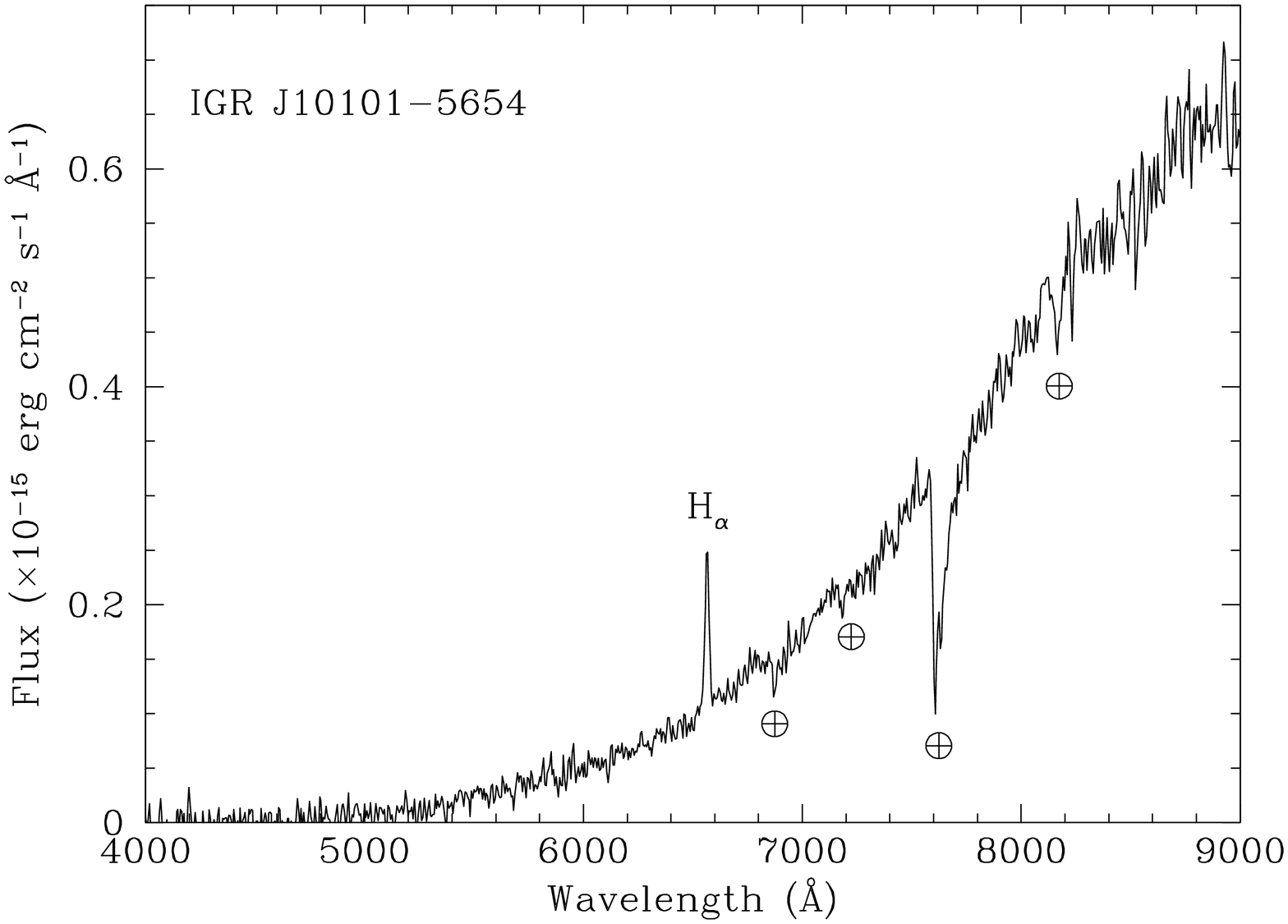,width=10.1cm,angle=270}}%}

\hspace{-.8cm} 
%\centering{ 
\mbox{\psfig{file=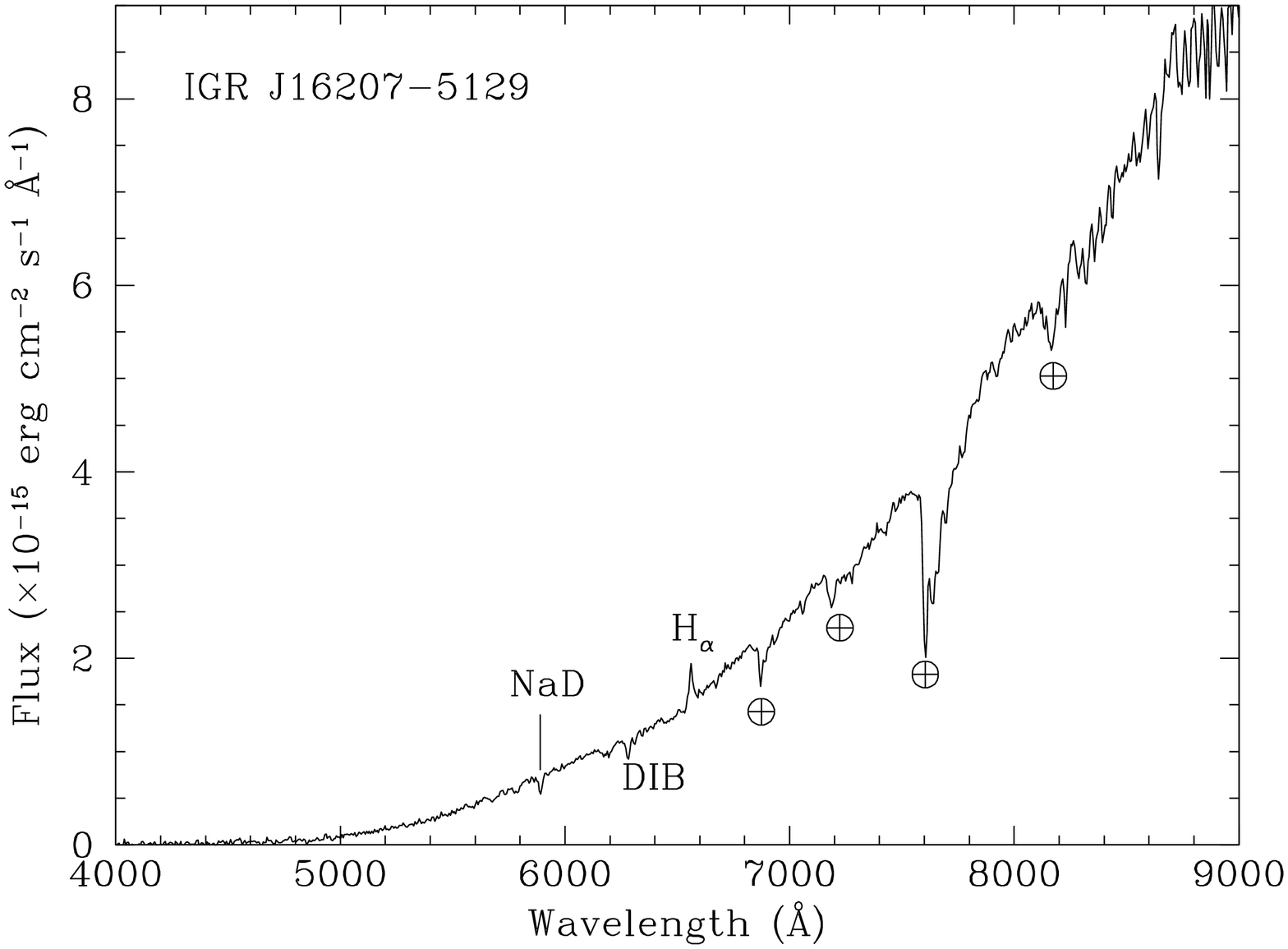,width=10.1cm,angle=270}}%} 
\hspace{-1cm} 
%\centering{ 
\mbox{\psfig{file=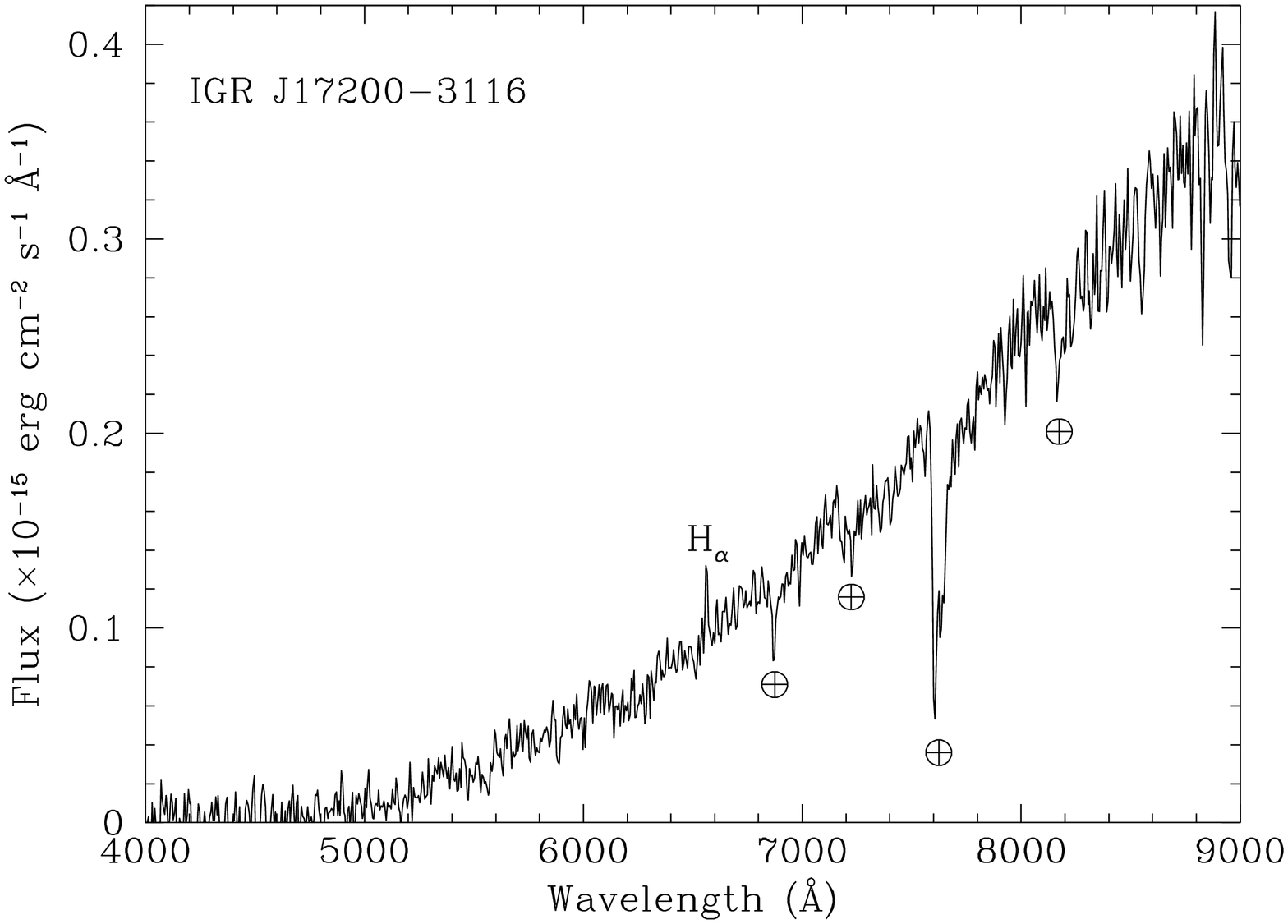,width=10.1cm,angle=270}}%} 
\vspace{-.5cm} 
\caption{Spectra (not corrected for the intervening Galactic absorption) 
of the optical counterparts of the HMXBs identified in this sample; 
all spectra were 
acquired with the 1.5m CTIO telescope. For each spectrum the main spectral 
features are labeled. The symbol $\oplus$ indicates atmospheric telluric 
absorption bands.}
%} 
%\end{center} 
\end{figure*}

\begin{figure*}%[th!]
%\begin{center}
\hspace{-.8cm}
%\centering
\mbox{\psfig{file=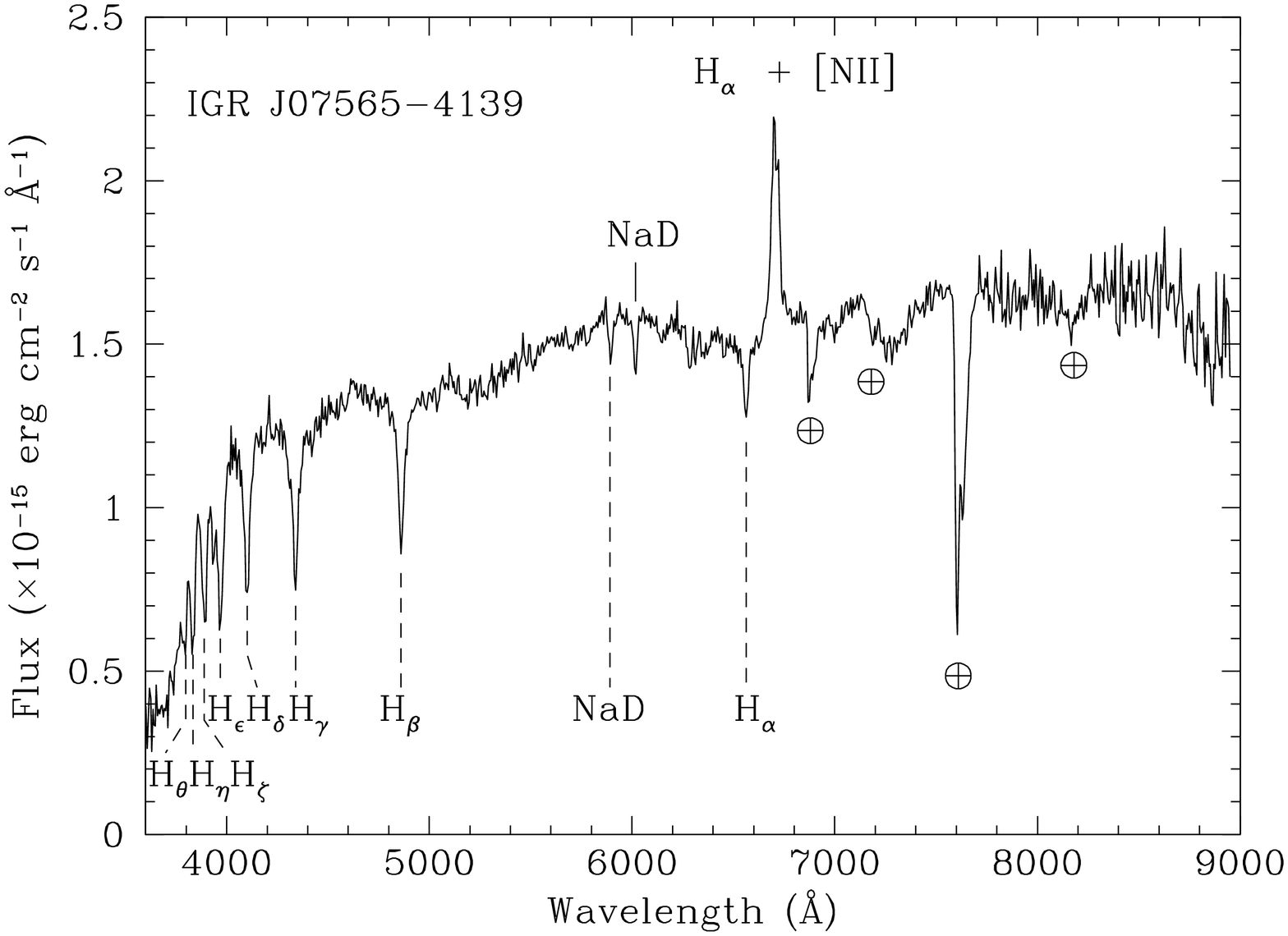,width=10.1cm,angle=270}}%}
\hspace{-1.0cm}
%\centering{
\mbox{\psfig{file=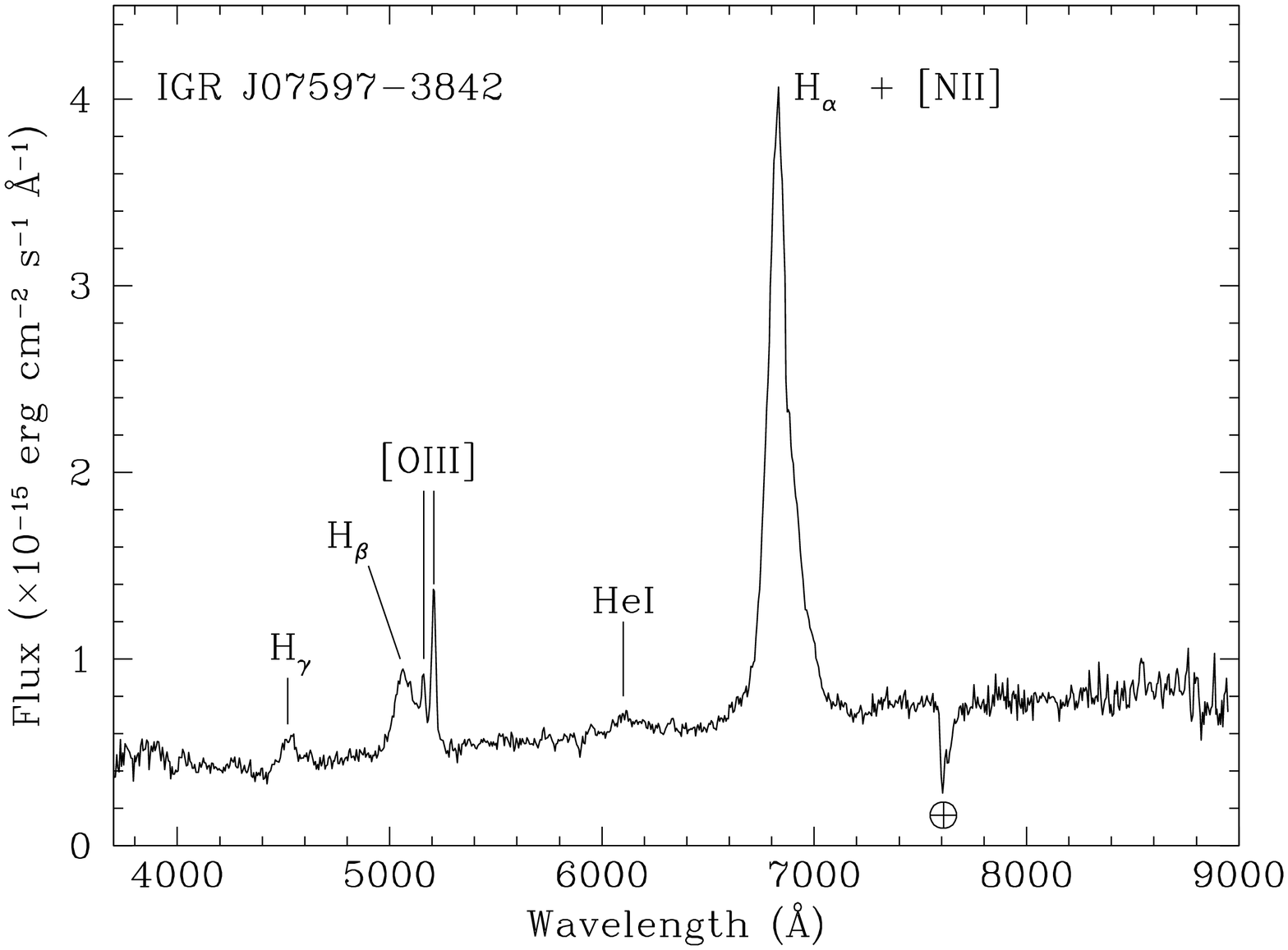,width=10.1cm,angle=270}}%}

\hspace{-.8cm}
%\centering
\mbox{\psfig{file=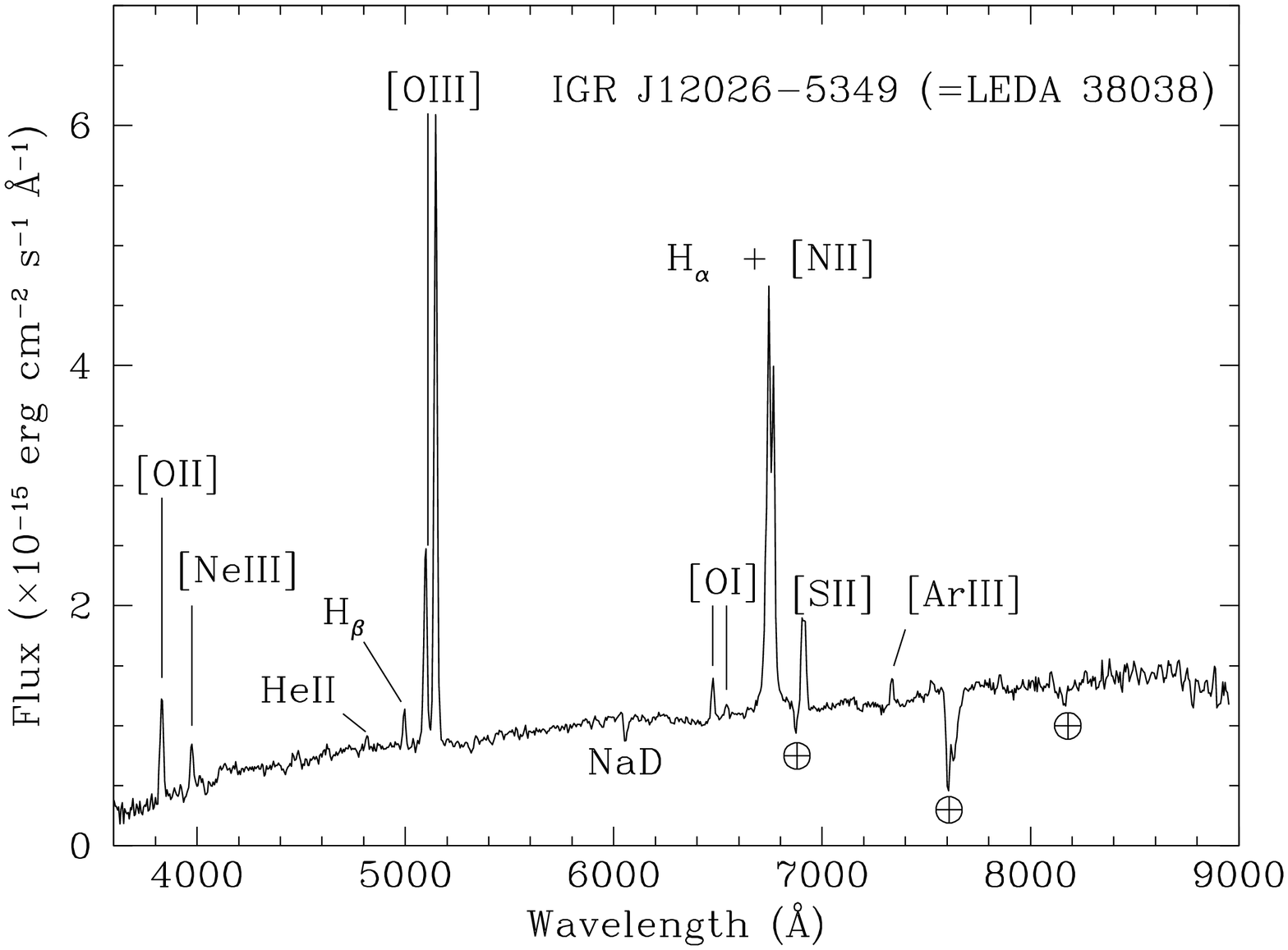,width=10.1cm,angle=270}}%}
\hspace{-1.0cm}
%\centering{
\mbox{\psfig{file=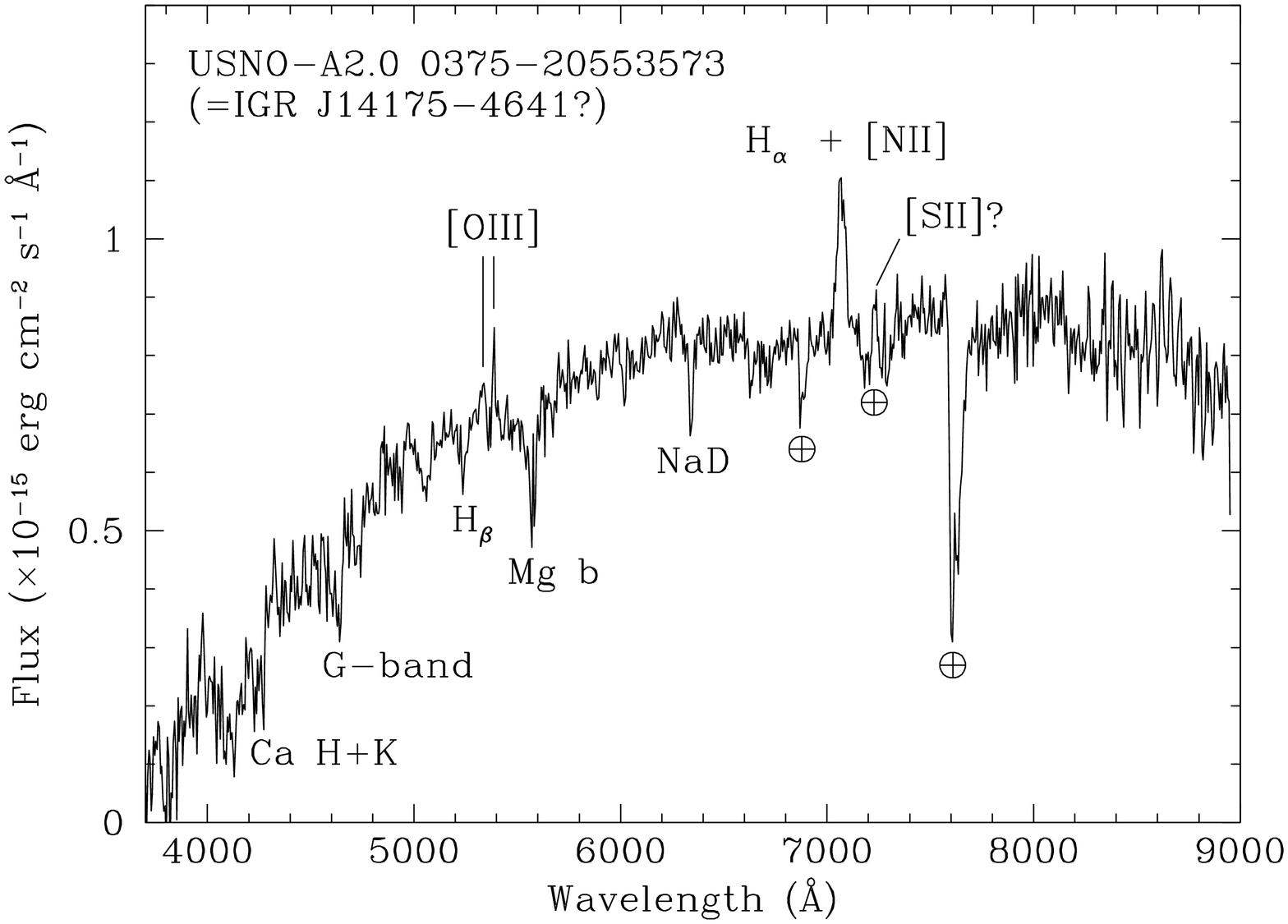,width=10.1cm,angle=270}}%}

\hspace{-.8cm}
%\centering{
\mbox{\psfig{file=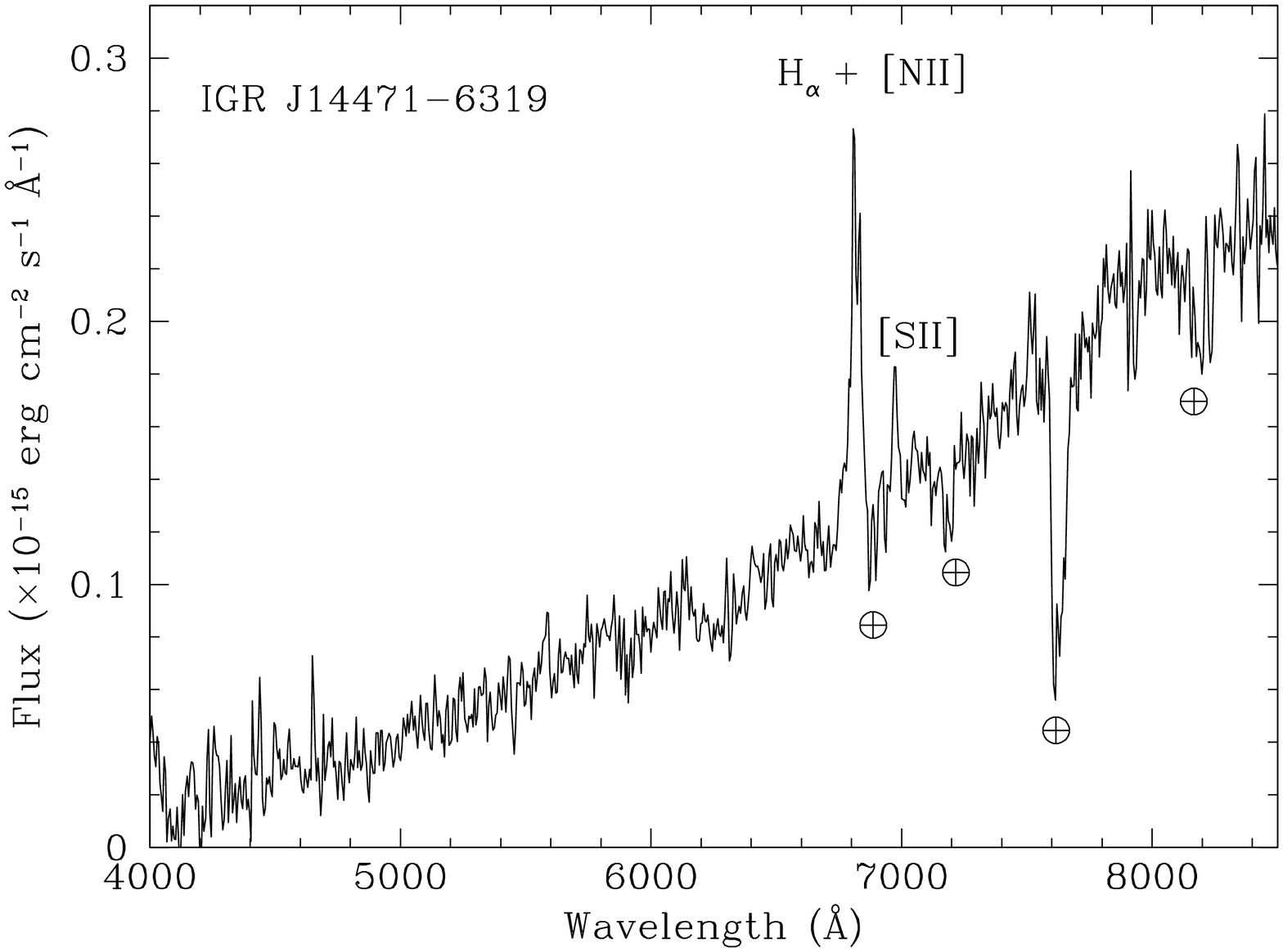,width=10.1cm,angle=270}}%}
\hspace{-1.0cm}
%\centering{
\mbox{\psfig{file=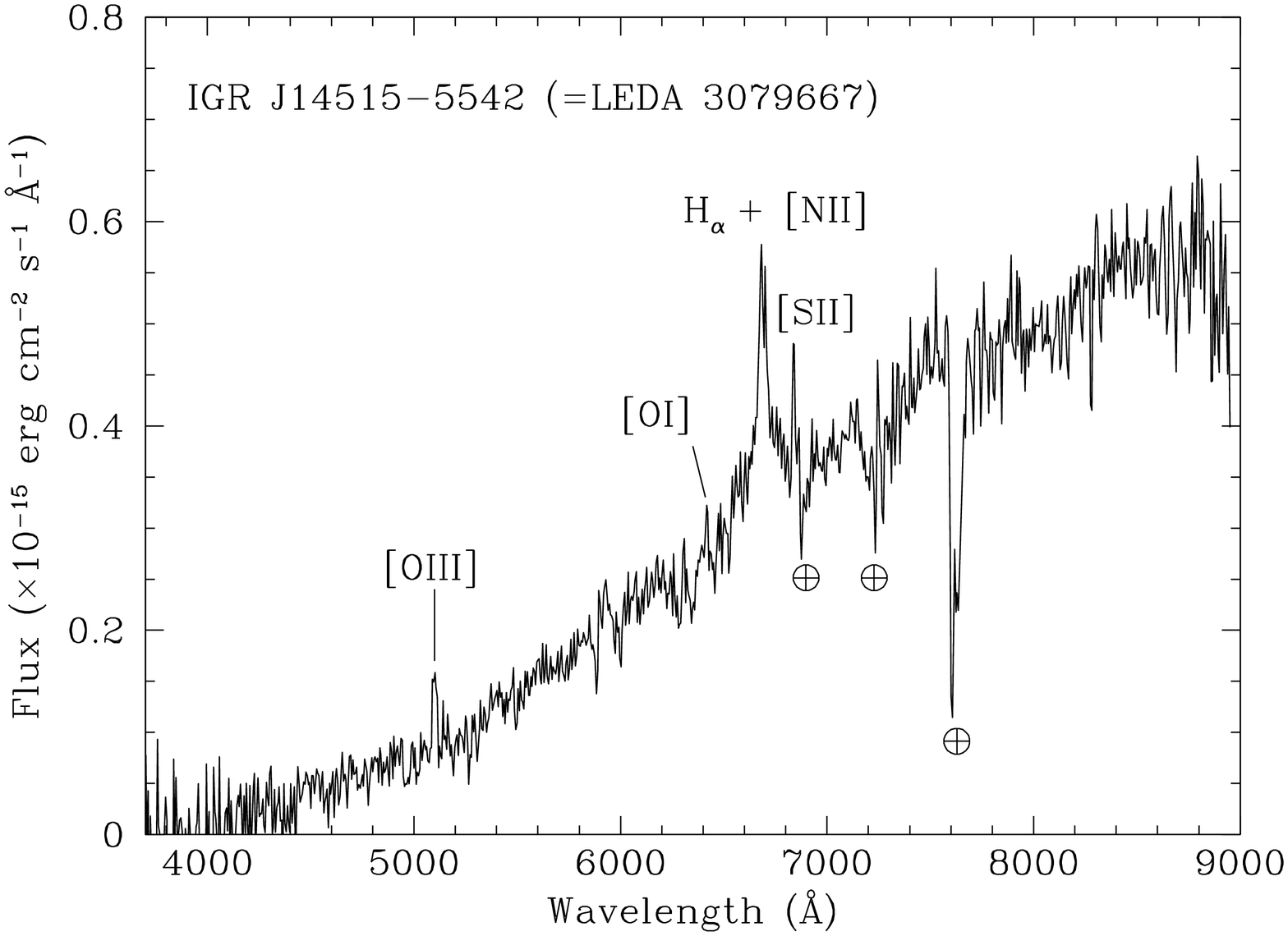,width=10.1cm,angle=270}}%}
\caption{Spectra (not corrected for the intervening Galactic absorption) 
of the optical counterparts of 6 AGNs identified in this sample; 
all spectra were 
acquired with the 1.5m CTIO telescope. For each spectrum the main spectral 
features are labeled. The symbol $\oplus$ indicates atmospheric telluric 
absorption bands. We remark that USNO-A2.0 object 0375$-$20553573 should 
be considered as the tentative, although likely, counterpart of IGR 
J14175$-$4641 (see text). The dashed lines in the spectrum of IGR 
J07565$-$4139 indicate absorption features at $z$ = 0, likely produced by 
an interloping Galactic star.}
%}
%\end{center}
\end{figure*}

\begin{figure*}%[th!]
%\addtocounter{figure}{-1}
%\begin{center}
%\hspace{.1cm}
\hspace{-.8cm}
%\centering{
\mbox{\psfig{file=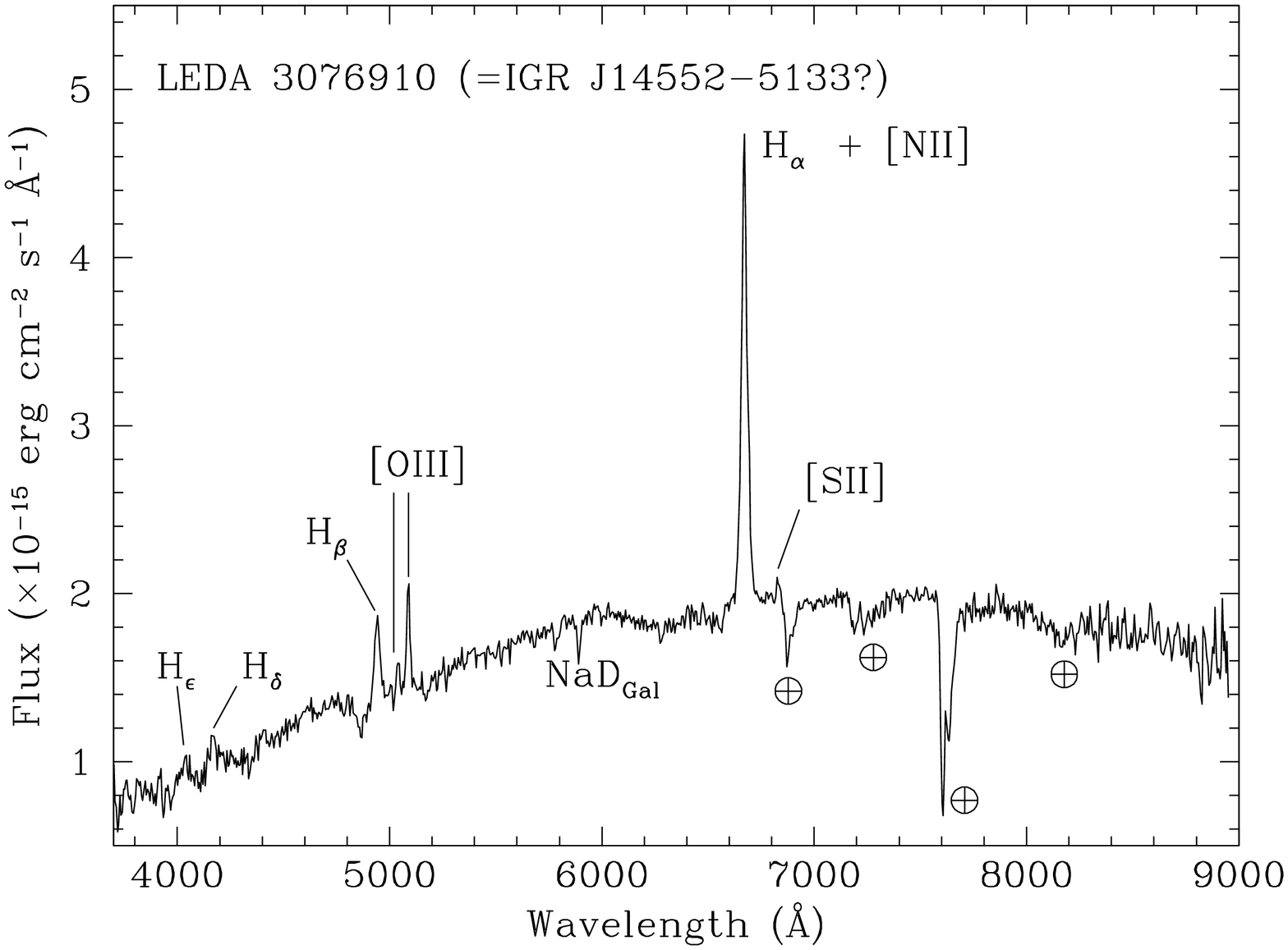,width=10.1cm,angle=270}}%}
\hspace{-1.0cm}
%\centering{
\mbox{\psfig{file=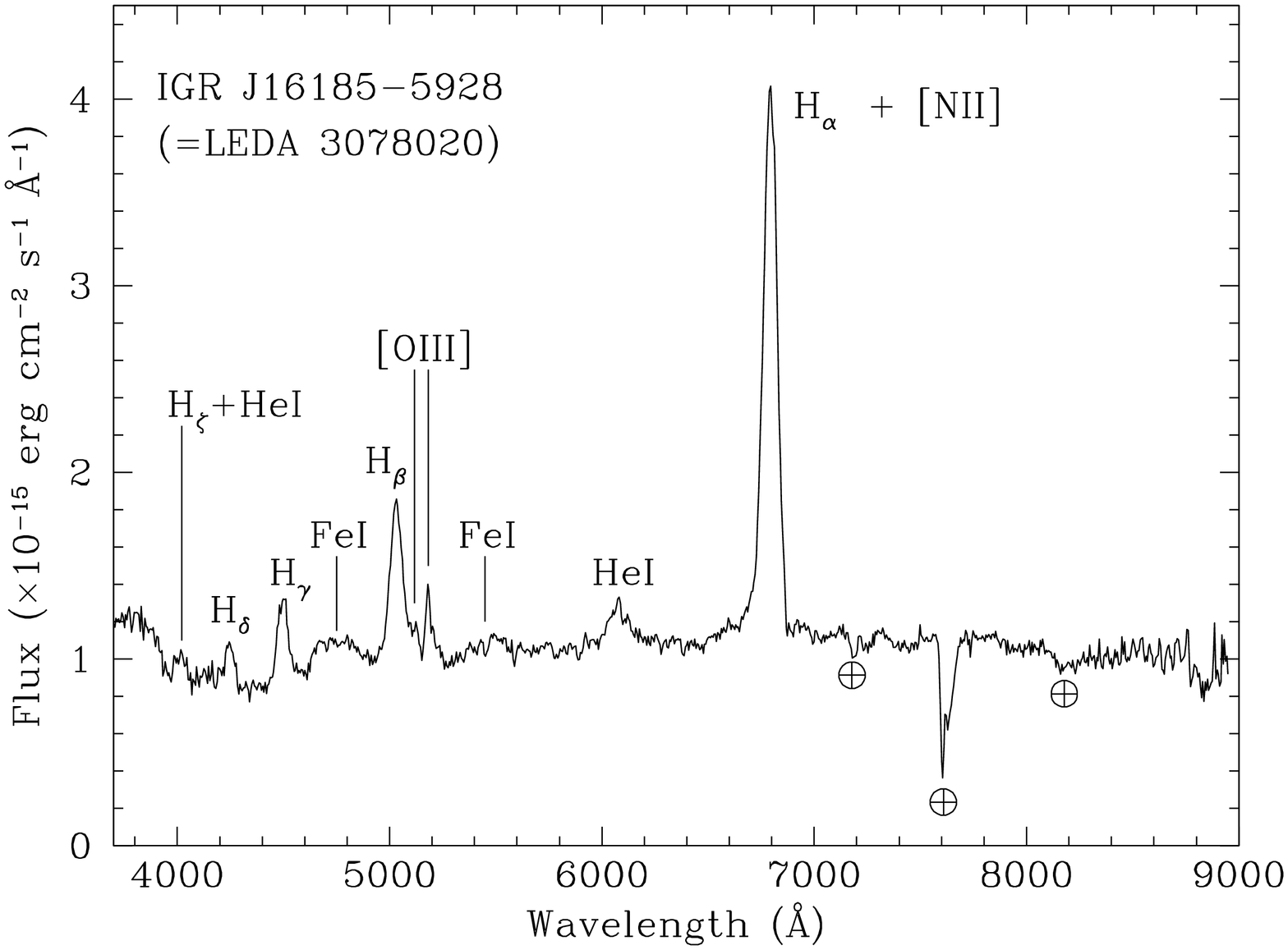,width=10.1cm,angle=270}}%}

\hspace{-.8cm}
%\centering{
\mbox{\psfig{file=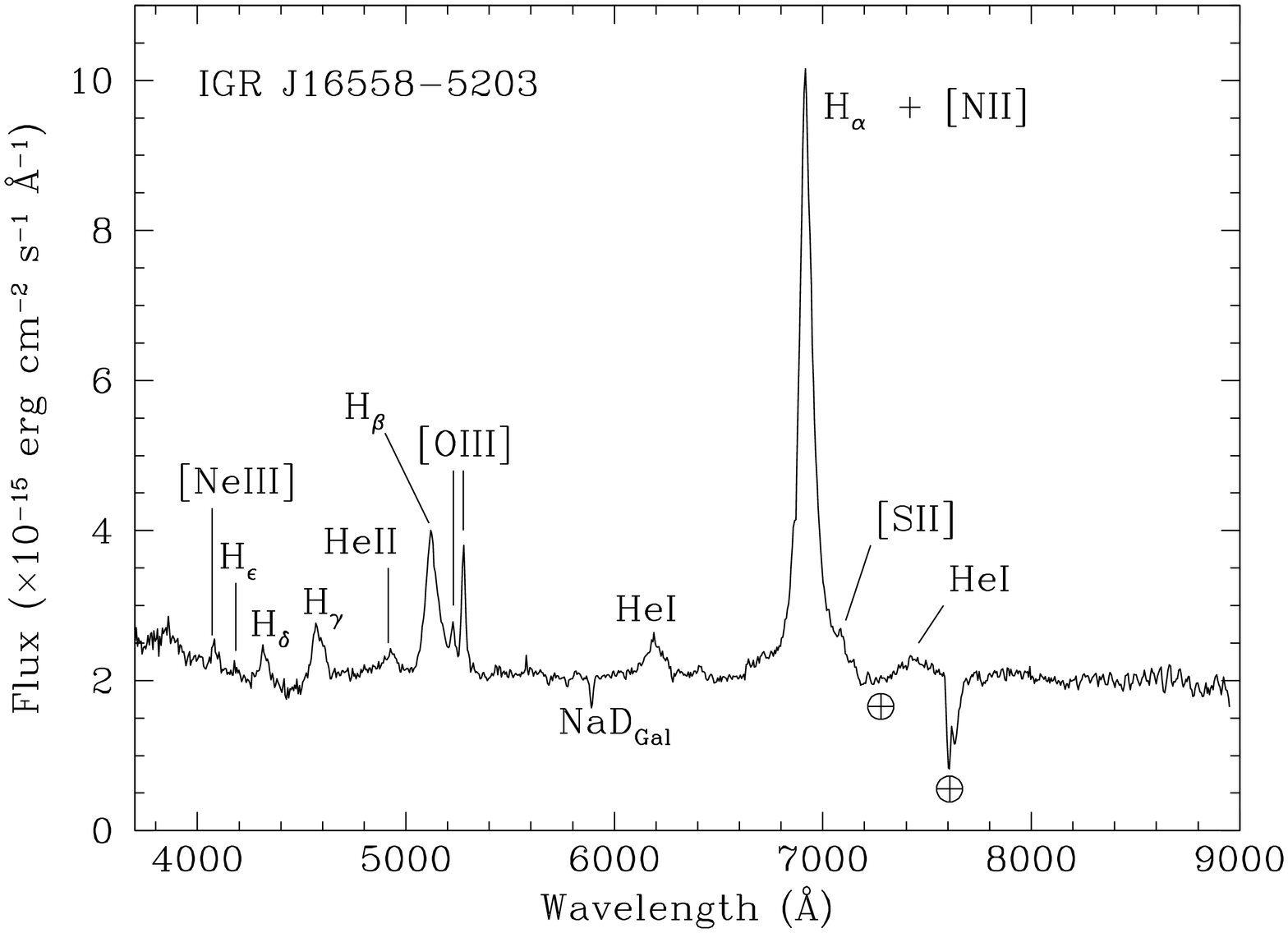,width=10.1cm,angle=270}}%}
\hspace{-1.0cm}
%\centering{
\mbox{\psfig{file=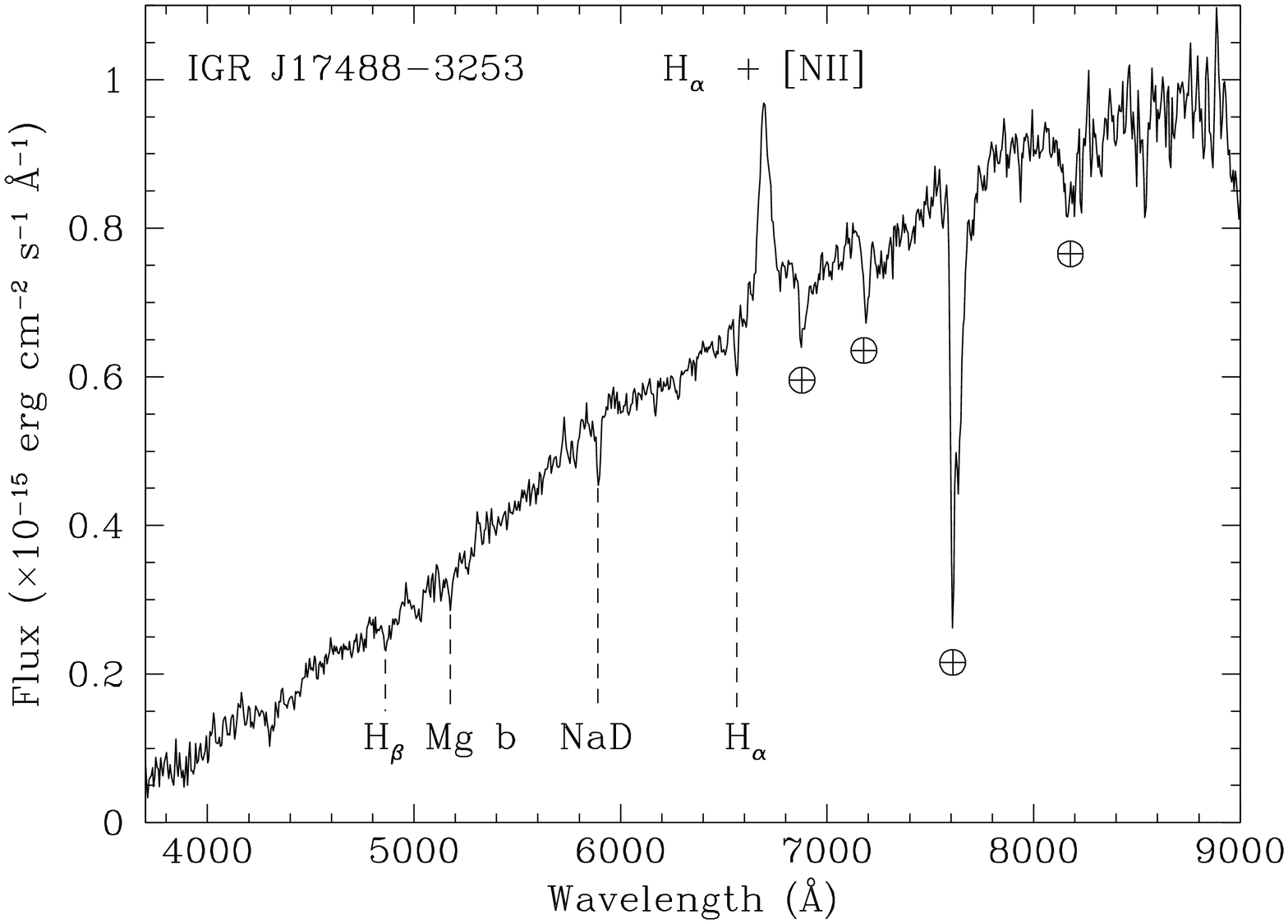,width=10.1cm,angle=270}}%}

\hspace{-.8cm}
%\centering{
\mbox{\psfig{file=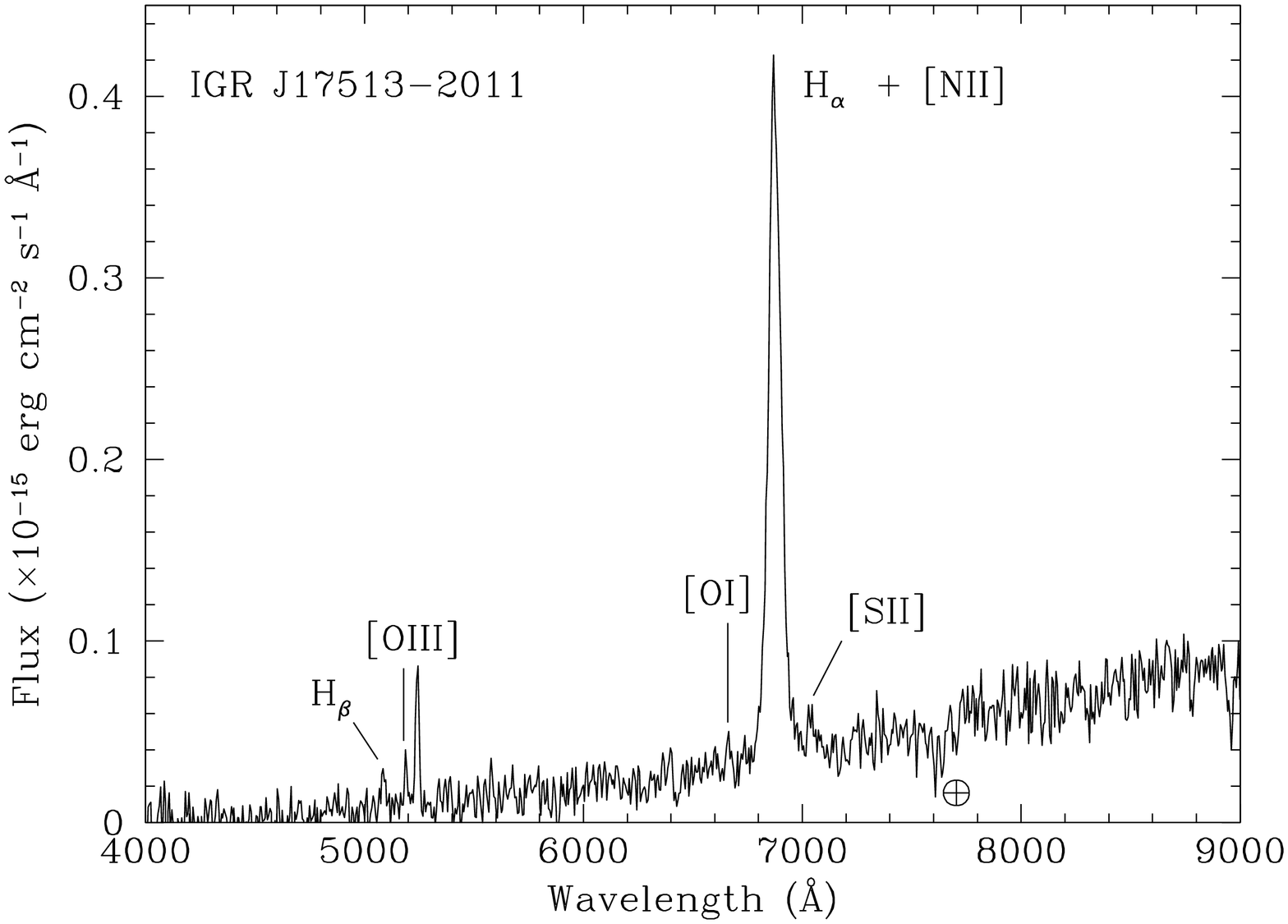,width=10.1cm,angle=270}}%}
\hspace{-1.0cm}
%\centering{
\mbox{\psfig{file=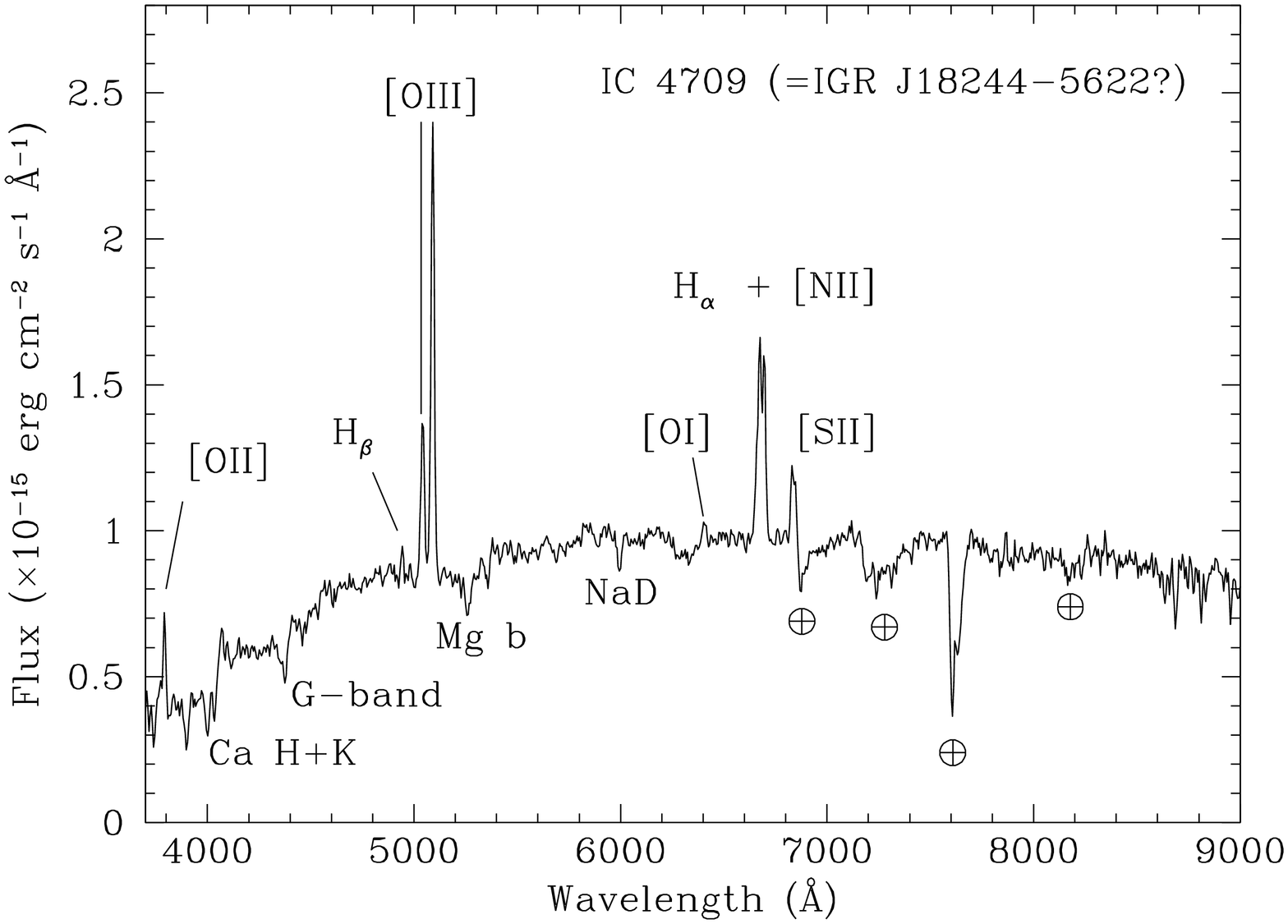,width=10.1cm,angle=270}}%}
\caption{
As Fig. 5, but for the remaining 6 AGNs identified in this sample.
We remark that 
galaxies LEDA 3076910 and IC 4709 should be considered as tentative, 
although likely, counterparts of IGR J14552$-$5133 and IGR J18244$-$5622, 
respectively (see text). The dashed lines in the spectrum of IGR
J17488$-$3253 indicate absorption features at $z$ = 0, likely produced by
an interloping Galactic star.}
%\end{center}
\end{figure*}

Five objects of our sample (XSS J12270$-$4859, IGR J14536$-$5522, IGR 
J15094$-$6649, IGR J16167$-$4957 and IGR J17195$-$4100) were identified as 
CVs through the appearance of their optical spectra (Fig. 3). All of them 
show Balmer emissions up to at least H$_\delta$, as well as several He 
{\sc i} and He {\sc ii} lines in emission. All of the detected lines are 
consistent with being at $z$ = 0, indicating that these objects belong to 
our Galaxy.

The {\it ROSAT} error circle of IGR J15094$-$6649 (see Fig. 2, upper 
middle panel) actually contains two relatively bright ($R <$ 18) objects: 
simultaneous optical spectroscopy shows that the true counterpart is the 
brighter one (the coordinates of which are reported in Table 2), while 
the other is a normal Galactic star with no peculiarities, thus 
unrelated with the {\it INTEGRAL} hard X--ray source.

The main spectral characteristics for each object, and the parameters 
which can be inferred from the optical and X--ray observational data, are 
listed in Table 3. The facts that, in the spectra of all these 5 objects, 
the Balmer decrement clearly appears negative, the He{\sc 
ii}$\lambda$4686/H$_\beta$ Equivalent Width (EW) ratio is $\ga$0.5 and the 
EWs of He {\sc ii} and H$_\beta$ are around (or larger than) 10 
\AA~indicate that these sources are magnetic CVs belonging to the 
Intermediate Polar (IP) subclass (see Warner 1995 and references therein). 
Their optical spectra, moreover, closely resemble that of another CV 
detected with {\it INTEGRAL}, IGR J00234+6141 (den Hartog et al. 2006; 
Halpern \& Mirabal 2006; Paper IV), which was classified as an IP as well 
(Paper IV; Bikmaev et al. 2006).

The large ($>$20 \AA) EWs observed in the He {\sc ii} $\lambda$4686 and 
H$_\beta$ lines of the spectrum of the optical counterpart of IGR 
J14536$-$5522 may indicate an even stronger magnetic field of the
accreting white dwarf harboured in this system. This fact may suggest 
that this object is perhaps a Polar, rather than an IP, CV (Warner 1995).

Given that, with the exception of IGR J16167$-$4957, the observed 
H$_\alpha$/H$_\beta$ flux ratio is always $<$1.5, we assumed $A_V$ = 0 
mag for the computation of the distance to each CV (see Table 3). 
For IGR J16167$-$4957 we instead considered $A_V$ = 1.3 mag, inferred by 
Tomsick et al. (2006) from a multiwavelength study of this source. 
The X--ray luminosities for the various objects were then computed using 
the fluxes reported in Bird et al. (2006), Kuiper et al. (2006) and 
Tomsick et al. (2006).

\begin{table*}%[t!]
\caption[]{Synoptic table containing the main results concerning the 5 CVs
discovered in the present sample of {\it INTEGRAL} sources. EWs are expressed 
in \AA, line fluxes are
in units of 10$^{-14}$ erg cm$^{-2}$ s$^{-1}$, whereas X--ray luminosities
are in units of 10$^{32}$ erg s$^{-1}$ and the reference band (between brackets)
is expressed in keV.}
\begin{center}
\begin{tabular}{lccccccccr}
\noalign{\smallskip}
\hline
\hline
\noalign{\smallskip}
\multicolumn{1}{c}{Object} & \multicolumn{2}{c}{H$_\alpha$} & 
\multicolumn{2}{c}{H$_\beta$} & \multicolumn{2}{c}{He {\sc ii} $\lambda$4686} & 
$R$ & $d$ & \multicolumn{1}{c}{$L_{\rm X}$} \\
\cline{2-7}
\noalign{\smallskip} 
 & EW & Flux & EW & Flux & EW & Flux & mag & (pc) & \\

\noalign{\smallskip}
\hline
\noalign{\smallskip}

XSS J12270$-$4859 & 10.2$\pm$0.8 & 0.54$\pm$0.04 & 11.6$\pm$1.7 & 0.75$\pm$0.11 & 12.4$\pm$1.9 & 0.83$\pm$0.12 & 
15.7 & $\sim$220 & 0.11 (0.1--2.4) \\
 & & & & & & & & & 1.4 (17--60) \\ 

 & & & & & & & & & \\ 

IGR J14536$-$5522 & 29$\pm$1 & 8.6$\pm$0.3 & 30.4$\pm$1.5 & 10.0$\pm$0.5 & 20.3$\pm$1.4 & 5.2$\pm$0.4 &
15.4 & $\sim$190 & 0.15 (0.1--2.4) \\
 & & & & & & & & & 1.3 (20--65) \\

 & & & & & & & & & \\ 

IGR J15094$-$6649 & 45$\pm$2 & 11.1$\pm$0.6 & 19.6$\pm$1.0 & 7.9$\pm$0.4 & 12.8$\pm$0.6 & 5.5$\pm$0.3 &
14.7 & $\sim$140 & 0.027 (0.1--2.4) \\
 & & & & & & & & & 0.35 (17--60) \\

 & & & & & & & & & \\ 

IGR J16167$-$4957 & 39$\pm$3 & 2.46$\pm$0.17 & 11.2$\pm$1.1 & 0.84$\pm$0.08 & 10.5$\pm$1.1 & 0.79$\pm$0.08 &
16.2 & $\sim$170 & 1.5 (0.5--10) \\
 & & & & & & & & & 0.55 (20--40) \\ 
 & & & & & & & & & $<$0.33 (40--100) \\ 

 & & & & & & & & & \\ 

IGR J17195$-$4100 & 85$\pm$3 & 16.7$\pm$0.5 & 41$\pm$2 & 10.6$\pm$0.5 & 18.5$\pm$0.9 & 5.5$\pm$0.3 &
14.3 & $\sim$110 & 0.36 (0.5--10) \\
 & & & & & & & & & 0.55 (20--100) \\ 

\noalign{\smallskip} 
\hline
\hline
\noalign{\smallskip} 
\end{tabular} 
\end{center}
\end{table*}

\subsection{HMXBs}

\begin{table*}%[t!]
\caption[]{Synoptic table containing the main results concerning the 4 HMXBs
discovered in the present sample of {\it INTEGRAL} sources. EWs are expressed 
in \AA, line fluxes are
in units of 10$^{-14}$ erg cm$^{-2}$ s$^{-1}$, whereas X--ray luminosities
are in units of 10$^{36}$ erg s$^{-1}$ and the reference band (between brackets) 
is expressed in keV.}
\begin{center}
\begin{tabular}{lccccccr}
\noalign{\smallskip}
\hline
\hline
\noalign{\smallskip}
\multicolumn{1}{c}{Object} & \multicolumn{2}{c}{H$_\alpha$} & 
Optical & $A_V$ & $d$ & Spectral & \multicolumn{1}{c}{$L_{\rm X}$} \\
\cline{2-3}
\noalign{\smallskip} 
 & EW & Flux & mag. & (mag) & (kpc) & type & \\

\noalign{\smallskip}
\hline
\noalign{\smallskip}

IGR J05007$-$7047 & 8.8$\pm$0.4  & 2.3$\pm$0.1 & 14.8 ($V$) & 0.38 & 
50 & B2\,III & 0.91$^{\rm a}$ (0.5--8) \\
 & & & & & & & 3.6$^{\rm a}$ (17--60) \\

 & & & & & & & \\ 

IGR J10101$-$5654 & 34$\pm$4 & 0.33$\pm$0.04 & --- & --- & --- & early giant & 
\multicolumn{1}{c}{---} \\ 

 & & & & & & & \\ 

IGR J16207$-$5129 & 5.0$\pm$0.3  & 0.76$\pm$0.04 & $\sim$15.6 ($R$) & 10.8 & $\sim$4.6$^{\rm b}$ 
& early supergiant & 0.11 (0.3--10) \\
 & & & & & & & 0.12 (20--100) \\

 & & & & & & & \\ 

IGR J17200$-$3116 & 5.5$\pm$1.1  & 0.05$\pm$0.01 & --- & --- & --- & --- & 
\multicolumn{1}{c}{---} \\

\noalign{\smallskip} 
\hline
\noalign{\smallskip} 
\multicolumn{8}{l}{$^{\rm a}$: luminosity estimate from Sazonov et al. (2005)} \\
\multicolumn{8}{l}{$^{\rm b}$: assuming $(V-R)_0 \sim$ $-$0.1} \\
\noalign{\smallskip} 
\hline
\hline
\end{tabular} 
\end{center} 
\end{table*}

We identify 4 of the {\it INTEGRAL} sources of our sample as HMXBs by 
their overall spectral appearance (see Fig. 4), which is typical of this 
class of objects (see e.g. Paper III), with narrow H$_\alpha$ emission at 
a wavelength consistent with that of the laboratory restframe, 
superimposed on an intrinsically blue continuum with Balmer absorptions. 
In three cases out of four (IGR J10101$-$5654, IGR J16207$-$5129 and IGR 
J17200$-$3116), however, the stellar continuum appears strongly reddened 
and almost undetected blueward of 5000 \AA, implying the presence of 
substantial interstellar dust along the line of sight. This also is 
quite typical of HMXBs detected with {\it INTEGRAL} (e.g., 2RXP 
J130159.6$-$635806; Paper III) and indicates that these objects are 
relatively far from Earth.

In the case of IGR J05007$-$7047 we detect the H$_\alpha$ line in 
emission, and the rest of the Balmer series (up to H$_\zeta$) in absorption,
all redshifted of $\sim$6 \AA~with respect to the corresponding laboratory 
wavelengths. This is consistent with the redshift of the Large Magellanic 
Cloud (LMC; see Cusumano et al. 1998 and references therein); thus, we 
confirm the hypothesis put forward by Sazonov et al. (2005) and G\"otz et 
al. (2006) that this hard X--ray source is indeed a HMXB belonging to the 
LMC.

We here note that the {\it Swift} error circle of IGR J10101$-$5654 (Fig. 
1, upper middle panel) marginally encompasess a relatively bright object: 
in this case also our optical spectroscopy, acquired simultaneously with 
that of the true counterpart, shows that it is a normal mid-type Galactic 
star.

Table 4 collects the relevant optical spectral information on these 4 
sources, along with their main parameters inferred from the available 
optical and X--ray data. Luminosities for each object were calculated 
using the X--ray fluxes in Bird et al. (2006) and Tomsick et al. (2006). 

Concerning IGR J05007$-$7047, optical photometry was obtained from 
Massey (2002), and a distance of 50 kpc was assumed.

For IGR J16207$-$5129, thanks to the subarcsec {\it Chandra} position of 
Tomsick et al. (2006), we revise the identification given in Paper III: 
this object is still a HMXB, but is not connected at all with star HD 
146803 (which lies in the IBIS error box, but it is not consistent with
the {\it Chandra} soft X--ray position). Assuming $A_V$ = 10.8 (Tomsick 
et al. 2006), we determine for this source the parameter values reported 
in Table 4.

Unfortunately, due to the lack of reliable optical photometry for the 
optical counterparts of IGR J10101$-$5654 and IGR J17200$-$3116, no 
significant information concerning distance, spectral type and X--ray 
luminosity can be determined for these two objects. Nevertheless, the 
H$_\alpha$ EW we measure for IGR J10101$-$5654 appears too large for a 
supergiant secondary star (see Leitherer 1988). Thus, we infer that this 
HMXB hosts a secondary of intermediate luminosity class.

\subsection{AGNs}

\begin{table*}[th!]
\caption[]{Synoptic table containing the main results concerning the 12 AGNs 
discovered in the present sample of {\it INTEGRAL} sources. Emission line
fluxes are reported both as observed and (between square brackets) corrected 
for the intervening Galactic absorption $E(B-V)_{\rm Gal}$ along the object
line of sight.
Line fluxes are in units of 10$^{-14}$ erg cm$^{-2}$ s$^{-1}$, whereas X--ray 
luminosities are in units of 10$^{43}$ erg s$^{-1}$ and the reference band 
(between round brackets) is expressed in keV. Errors and limits are at 
1$\sigma$ and 3$\sigma$ confidence levels, respectively. The typical error on 
the redshift measurement is $\pm$0.001.}
\begin{center}
\begin{tabular}{lcccccrccr}
\noalign{\smallskip}
\hline
\hline
\noalign{\smallskip}
\multicolumn{1}{c}{Object} & $F_{\rm H_\alpha}$ & $F_{\rm H_\beta}$ &
$F_{\rm [OIII]}$ & Class & $z$ & 
\multicolumn{1}{c}{$D_L$} & \multicolumn{2}{c}{$E(B-V)$} & 
\multicolumn{1}{c}{$L_{\rm X}$} \\
\cline{8-9}
\noalign{\smallskip}
 & & & & & & (Mpc) & Gal. & AGN & \\
\noalign{\smallskip}
\hline
\noalign{\smallskip}

IGR J07565$-$4139 & 0.56$\pm$0.08 & $<$0.05 & $<$0.04 & Sy2 & 0.021 & 
98.4 & 0.77 & $>$1.4 & 0.41 (0.5--8) \\
 & [7.0$\pm$1.0] & [$<$0.6] & [$<$0.4] & & & & & & 1.7 (17--60) \\
 & & & & & & & & & 0.88 (20--40) \\
 & & & & & & & & & $<$8.7 (40--100) \\

 & & & & & & & & & \\

IGR J07597$-$3842 & * & 4.9$\pm$0.5 & 2.1$\pm$0.2 & Sy1.2 & 0.040 & 
190.1 & 0.81 & 0 & 3.5 (0.1--2) \\
 & * & [61$\pm$6] & [25$\pm$3] & & & & & & 15.3 (20--100) \\

 & & & & & & & & & \\

IGR J12026$-$5349 & 4.6$\pm$0.4 & 0.5$\pm$0.1 & 9.1$\pm$0.5 & Sy2 & 0.028 & 
131.4 & 0.21 & 1.0 & 1.4 (0.5--8) \\
 & [7.6$\pm$0.7] & [1.0$\pm$0.2] & [17.5$\pm$0.9] & & & & & & 6.8 (17--60) \\

 & & & & & & & & & \\

IGR J14175$-$4641 &  0.64$\pm$0.13 & in abs. & 0.18$\pm$0.04 & Sy2 & 0.076 & 
370.6 & 0.11 & --- & 27.1 (17--60) \\
 & [0.79$\pm$0.16] & $''$ & [0.24$\pm$0.05] & & & & & & \\

 & & & & & & & & & \\

IGR J14471$-$6319 & 0.17$\pm$0.02 & $<$0.02 & $<$0.02 & Sy2 & 0.038 & 
180.4 & 1.31 & $>$0.58 & 0.044 (0.1--2.4) \\
 & [7.6$\pm$1.1] & [$<$1.5] & [$<$1.1] & & & & & & 4.8 (17--60) \\

 & & & & & & & & & \\

IGR J14515$-$5542 & 0.45$\pm$0.09 & $<$0.03 & 0.18$\pm$0.04 & Sy2 & 0.018 & 
84.2 & 1.13 & --- & 0.11 (0.1--2.4) \\
 & [2.9$\pm$0.6] & [$<$3.0] & [6.4$\pm$1.6] & & & & & & 1.8 (20--65) \\

 & & & & & & & & & \\

IGR J14552$-$5133 & * & 1.81$\pm$0.12 & 1.26$\pm$0.07 & NL Sy1 & 0.016 & 
74.7 & 0.67 & 0 & 0.079 (0.1--2.4) \\
 & * & [14.8$\pm$1.0] & [10.2$\pm$0.7] & & & & & & 0.92 (17--60) \\

 & & & & & & & & & \\

IGR J16185$-$5928 & * & 6.0$\pm$0.3 & 1.21$\pm$0.15 & NL Sy1 & 0.035 & 
165.8 & 0.32 & 0 & 0.50 (0.1--2.4) \\
 & * & [16.0$\pm$0.8] & [3.2$\pm$0.4] & & & & & & 5.4 (17--60) \\

 & & & & & & & & & \\

IGR J16558$-$5203 & * & 14.6$\pm$0.8 & 3.9$\pm$0.3 & Sy1.2 & 0.054 & 
259.3 & 0.44 & 0 & 3.0 (0.1--2.4) \\
 & * & [53$\pm$3] & [14.7$\pm$1.0] & & & & & & 27.4 (20--100) \\

 & & & & & & & & & \\

IGR J17488$-$3253 & * & $<$0.08 & $<$0.1 & Sy1 & 0.020 & 
93.7 & 1.63 & 0 & 0.098 (0.1--2.4) \\
 & * & [$<$10] & [$<$16] & & & & & & 5.1 (20--100) \\

 & & & & & & & & & \\

IGR J17513$-$2011 & * & 0.047$\pm$0.014 & 0.15$\pm$0.02 & Sy1.9 & 0.047 & 
224.5 & 1.28 & 0 & 0.24 (0.1--2.4) \\
 & * & [2.3$\pm$0.7] & [6.9$\pm$0.7] & & & & & & 23.2 (20--100)\\

 & & & & & & & & & \\

IGR J18244$-$5622 & 1.16$\pm$0.09 & 0.14$\pm$0.05 & 3.05$\pm$0.16 & Sy2 & 0.017 & 
79.4 & 0.09 & 1.17 & 0.83 (2--10) \\
 & [1.41$\pm$0.11] & [0.16$\pm$0.05] & [3.9$\pm$0.2] & & & & & & \\

\noalign{\smallskip} 
\hline
\noalign{\smallskip} 
\multicolumn{8}{l}{$^*$: heavily blended with [N {\sc ii}] lines} \\
\noalign{\smallskip} 
\hline
\hline
\end{tabular} 
\end{center} 
\end{table*}

The remaining 12 objects of our sample show optical spectra which are 
dominated by redshifted broad and/or narrow emission lines typical of 
AGNs. We classify half of this subsample (6 objects) as Seyfert 2 
galaxies; the other half is made of Seyfert 1 galaxies.
Of these latter ones, 2 are classified as Seyfert 1.2, one as Seyfert 1.9 
and two as Narrow-Line (NL) Seyfert 1; for the case of IGR J17488$-$3253
only a general Seyfert 1 classification can be given due to the lower 
quality of the spectrum (see Fig. 6, central right panel).
The main observed and inferred parameters for each object are reported 
in Table 5. We assumed a null local absorption for Seyfert 1 AGNs.
In the table, X--ray luminosities were computed using the fluxes reported 
in Bird et al. (2006), Kuiper et al. (2006), Sazonov et al. (2005) and 
Revnivtsev et al. (2006a,b), and using the {\it ROSAT} countrates (Voges 
et al. 1999, 2000; {\it ROSAT} Team 2000).

One can note that the {\it ROSAT} error circles of IGR J14471$-$6319 (Fig. 
1, central middle panel) and IGR J14552$-$5133 (Fig. 2, upper left panel) 
include other objects, besides the ones we identify here as their optical 
counterparts. Optical spectroscopy demonstrates that these other objects 
are normal Galactic stars, therefore unrelated with the high energy 
emission detected by {\it INTEGRAL}.

We recall that, as mentioned in Sect. 2, the putative optical counterparts 
of hard X--ray sources IGR J14175$-$4641, IGR J14552$-$5133 and IGR 
J18244$-$5622 (Revnivtsev et al. 2006a,b) were not chosen via soft X--ray 
catalogues cross-correlation. We also remark that the first and the third 
of these cases completely lack a catalogued arcsec-sized X--ray counterpart 
position, while the second one contains three {\it ROSAT} sources within 
the IBIS error circle (with that associated with the selected 
putative optical counterpart being the closest to the centre of the IBIS 
error box). Therefore, the counterparts we propose for them, although 
likely, need confirmation through pointed soft X--ray observations with 
satellites affording arcsecond localizations (such as {\it Chandra}, {\it 
XMM-Newton} or {\it Swift}). Keeping in mind all of the above, from now on 
we will nevertheless consider as true the optical/hard X--ray association 
regarding these three sources.

Going into details concerning some objects, for two Seyfert 2 AGNs the 
available soft X--ray information (Sazonov et al. 2005; Revnivtsev et al 
2006b), together with the local absorption estimate obtained from the 
optical spectra (see Table 5), allows us to determine their Compton regime 
(see Bassani et al. 1999 for details). Using the [O{\sc iii}] $\lambda$5007 
line fluxes corrected for the total (Galactic plus local to the AGN) 
absorption, we find that IGR J12026$-$5349 has a [O{\sc iii}]/(2--10 keV) 
flux ratio of 0.8, indicating that it is a Compton thick source, whereas 
this same parameter is 5.0 for IGR J18244$-$5622, telling 
us that this is a Seyfert 2 AGN in the Compton thin regime.
In this respect, we remark that observations with satellites sensitive
to the 2--10 keV emission are needed for the other sources in Table 5 
lacking information in this X--ray band in order to definitely determine 
their Compton nature. For some of them, an analysis of the {\it Swift}
data is now underway (Landi et al., in preparation).

It is also noticeable that the spectra of IGR J07565$-$4139 and, to a 
lesser extent, IGR J17488$-$3253 appear as `hybrid', i.e. with features of 
both a Galactic star and of an AGN (the former ones are indicated with 
dashed lines in Fig. 5, upper left panel and in Fig. 6, central right 
panel). In this sense, they are evidently similar to the case of IGR 
J20247+5058 (Paper I), and can be interpreted as the chance superposition 
of a nearby star and a background AGN. High-resolution imaging is thus 
desirable to disentangle the Galactic star contribution for these two 
objects.

Moreover, as already mentioned, we tentatively classify IGR J14552$-$5133 
and IGR J16185$-$5928 as NL Seyfert 1 AGNs following the criteria of 
Osterbrock \& Pogge (1985), although the Fe {\sc ii} bumps are not readily 
detected in IGR J14552$-$5133 and the Full Width at Half Maximum (FWHM) of 
H$_\beta$ emission in IGR J16185$-$5928 is $\sim$4000 km s$^{-1}$, i.e. 
twice as the maximum expected for a NL Seyfert 1 AGN.

To conclude this Section, following Wu et al. (2004) and Kaspi et al. 
(2000), we can compute an estimate of the mass of the central black hole 
in 5 of the 6 objects classified as Seyfert 1 AGN (this procedure could 
not be applied to IGR J17488$-$3253 as no H$_\beta$ emission was 
detected). The broad-line region (BLR) gas velocities $v_{\rm BLR}$ 
(measured from the H$_\beta$ emission line FWHM) and the corresponding 
black hole masses for these 5 cases are reported in Table 6.

\subsection{Statistical considerations}

We can now briefly recover the statistical approach made in Paper II, 
updating the numbers presented there with recent discoveries (Papers III 
and IV; Masetti et al. 2006e,f; Halpern \& Gotthelf 2006; Negueruela \& 
Smith 2006) and with the sample of sources illustrated in the present 
work. It is now found that, presently, of the 54 {\it INTEGRAL} sources 
identified through optical spectroscopy, 22 (41\%) are X--ray binaries 
(with a large majority, i.e. more than 90\%, of HMXB), 24 (44\%) are AGNs 
(half of which were presented in this paper for the first time) and 8 
(15\%) are CVs (5 of which were shown here for the first time), with
at least 6 of them belonging to the IP subclass (see Paper IV and the 
present work).

One can compare, for instance, these numbers with those for the group of 
the 153 identified objects belonging to the largest catalogue of {\it 
INTEGRAL} sources published up to now, i.e., the 2$^{\rm nd}$ IBIS 
Galactic Plane Survey (Bird et al. 2006). In this survey we have 107 
(70\%) X--ray binaries (of which, only one third are HMXBs), 27 (18\%) 
AGNs and 8 (5\%) CVs, with at least 6 of them of magnetic nature
(IPs or Polars).

From these numbers, graphically reported in percentage terms in the 
histogram in Fig. 7, one can immediately see that the CV sample detected 
with {\it INTEGRAL} has been doubled thanks to the optical spectroscopy 
identification approach. This also stresses {\it INTEGRAL}'s sensitivity 
in detecting hard X--ray emission from this class of objects.
One of the reasons for this may be found in the fact that the bulk of the 
X--ray emission from magnetic CVs falls in the 20--40 keV band
(e.g., de Martino et al. 2004; Suleimanov et al. 2005), which is
the one in which {\it INTEGRAL} has the strongest sensitivity.
A thorough description of CVs detected by {\it INTEGRAL} up to now 
can be found in Barlow et al. (2006).

The number of {\it INTEGRAL}-detected AGNs also has nearly doubled 
thanks to these optical studies; besides, it is apparent that an important
fraction of the {\it INTEGRAL} sources identified by means of optical 
spectroscopy and lying on the Galactic Plane is composed of background 
AGNs. This once again underscores the extraordinary capabilities of 
{\it INTEGRAL} of piercing through the Zone of Avoidance of the Galaxy 
for the exploration of this part of the extragalactic sky.

As a final corollary, we would like here to stress the extreme 
effectiveness of the strategy of catalogues cross-correlation plus optical 
spectroscopy we are pursuing to securely pinpoint the actual nature of the 
X--ray sources detected with {\it INTEGRAL}: for instance, of the 56 
unidentified objects belonging to the 2$^{\rm nd}$ IBIS Galactic Plane 
Survey (Bird et al. 2006), this observational approach led to the 
discovery of the nature of 18 sources (10 AGNs, 6 HMXBs and 2 CVs), i.e., 
nearly one third of the total, 15 of which being reported in our Papers 
II-IV and in the present work. The corresponding source type percentages 
are represented as shaded areas in Fig. 7.

The lack of known accurate (up to $\sim$10 arcsec) soft X--ray position is 
the main cause of failure in this identification task; therefore, 
observations with high-resolution imaging X--ray satellites (such as {\it 
Chandra}, {\it XMM-Newton} and/or {\it Swift}) are of paramount importance 
for the continuation of this program aimed at identifying the nature of 
unknown {\it INTEGRAL} hard X--ray sources.

\begin{table}%[h!]
\caption[]{BLR gas velocities (in km s$^{-1}$) and 
central black hole masses (in units of 10$^7$ $M_\odot$) for 5
Seyfert 1 AGNs belonging to the sample presented in this paper.}
\begin{center}
%\vspace{-.3cm}
\begin{tabular}{lcc}
\noalign{\smallskip}
\hline
\hline
\noalign{\smallskip}
\multicolumn{1}{c}{Object} & $v_{\rm BLR}$ & $M_{\rm BH}$ \\
\noalign{\smallskip}
\hline
\noalign{\smallskip}

IGR J07597$-$3842 & 5500 & 20 \\ 
IGR J14552$-$5133 & 1600 & 0.2 \\
IGR J16185$-$5928 & 3500 & 2.8 \\
IGR J16558$-$5203 & 2900 & 7.8 \\
IGR J17513$-$2011 & 1100 & 0.1 \\

\noalign{\smallskip}
\hline
\hline
\noalign{\smallskip}
\end{tabular}
\end{center}
\end{table}

\begin{figure}%[t!]
\hspace{-0.5cm}
\psfig{file=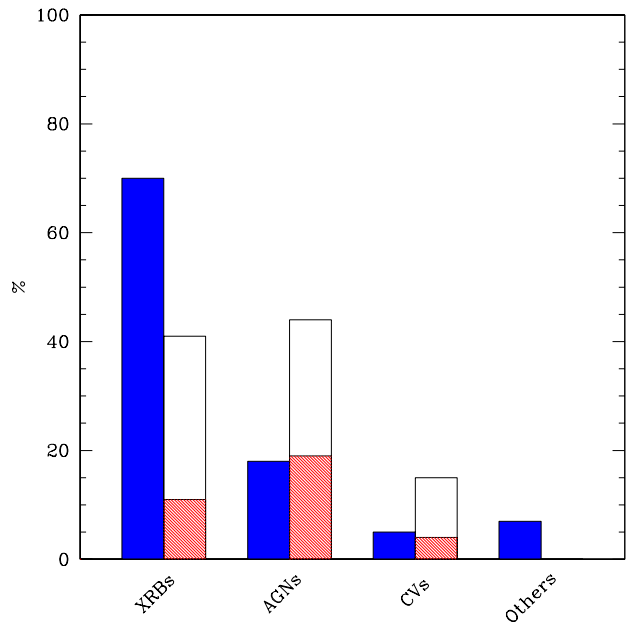,width=9.5cm}
\vspace{-0.7cm}
\caption{Histogram, subdivided into source types, showing the percentage 
of {\it INTEGRAL} objects of known nature and belonging to the 
2$^{\rm nd}$ IBIS Galactic Plane Survey (Bird et al. 2006; left-side, 
darker columns), and {\it INTEGRAL} sources from various surveys 
and identified through optical spectroscopy (right-side, lighter columns). 
The latter columns also show (as shaded areas) the percentage of sources
identified thanks to optical spectroscopy and which belong to 2$^{\rm nd}$ 
IBIS Galactic Plane Survey.}
\end{figure}

\section{Conclusions}

In our continuing work of identification of {\it INTEGRAL} sources by 
means of optical spectroscopy (Papers I-IV), we have identified and 
studied 21 southern hard X--ray objects of unknown nature by means of the 
1.5m CTIO telescope of Cerro Tololo (Chile).

We found that the selected sample is made of 12 AGNs (6 of which are of 
Seyfert 1 type and 6 are Seyfert 2 AGNs), 5 magnetic CVs and 4 HMXBs (one 
of which in the LMC). In terms of relative sizes of the three groups, we 
notice the absolute majority of AGNs in the sample, and a comparatively 
large fraction ($\sim$25\%) of CVs.

We recall that in three cases (IGR J14175$-$4641, IGR J14452$-$5133 and 
IGR J18244$-$5622), all identified as AGNs, only a tentative albeit likely 
optical counterpart was given because of the the lack of an univocal 
arcsecond-sized soft X--ray position. Thus, for them an observation with 
soft X--ray satellites affording arcsecond localizations (such as {\it 
Chandra}, {\it XMM-Newton} or {\it Swift}) is needed to confirm the 
proposed association.

The results presented in this work further indicate the capabilities of 
{\it INTEGRAL} to reveal not only high- and low-mass X-ray binaries, but 
also (if not mostly) extragalactic objects in the nearby Universe 
($z <$ 0.1) and magnetic dwarf novae.

\begin{acknowledgements}

We thank Claudio Aguilera and Arturo G\'omez for the daily and nightly 
assistance at the telescope, and the anonymous referee for useful 
remarks which helped us to improve the paper. 
This research has made use of the NASA Astrophysics 
Data System Abstract Service, of the NASA/IPAC Extragalactic Database 
(NED), and of the NASA/IPAC Infrared Science Archive, which are operated 
by the Jet Propulsion Laboratory, California Institute of Technology, 
under contract with the National Aeronautics and Space Administration. 
This publication makes use of data products from the Two Micron All 
Sky Survey (2MASS), which is a joint project of the University of 
Massachusetts and the Infrared Processing and Analysis Center/California 
Institute of Technology, funded by the National Aeronautics and Space 
Administration and the National Science Foundation.
This research has also made use of the SIMBAD database operated at CDS, 
Strasbourg, France, and of the HyperLeda catalogue operated at the 
Observatoire de Lyon, France.
The authors acknowledge the ASI and INAF financial support via grant 
No. 1/023/05/0.
NM thanks the Departamento de Astronom\'{i}a y Astrof\'{i}sica of the 
Pontificia Universidad Cat\'olica of Santiago (Chile) for the pleasant 
hospitality during the preparation of this paper.
LM's work is supported by the Fondap Center for Astrophysics grant 
No. 15010003.
\end{acknowledgements}


\begin{thebibliography}{}

\bibitem{} Barlow, E.J., Knigge, C., Bird, A.J., et al. 2006, MNRAS, in 
	press {\tt [astro-ph/0607473]}

\bibitem{} Bassani, L., Dadina, M., Maiolino, R., et al. 1999, ApJS, 121,
        473

\bibitem{} Bird, A.J., Barlow, E.J., Bassani, L., et al. 2006, ApJ, 636, 765

\bibitem{} Bikmaev, I.F., Revnivtsev, M.G., Burenin, R.A., \& Sunyaev, R.A. 
	2006, Astron. Lett., in press {\tt [astro-ph/0603715]}

\bibitem{} Cardelli, J.A., Clayton, G.C., \& Mathis, J.S. 1989, ApJ, 345,
        245

\bibitem{} Cusumano, G., Israel, G.L., Mannucci, F., et al. 1998, A\&A, 
	337, 772

\bibitem{} den Hartog, P.R., Hermsen, W., Kuiper, L., et al. 2006, A\&A,
	451, 587

\bibitem{} de Martino, D., Matt, G., Belloni, T., Haberl, F., \& Mukai, K.
        2004, A\&A, 415, 1009

\bibitem{} Dean, A.J., Bazzano, A., Hill, A.B., et al. 2005, A\&A, 443, 485

\bibitem{} G\"otz, D., Mereghetti, S., Merlini, D., Sidoli, L., \& Belloni, 
	T. 2006, A\&A, 448, 873

\bibitem{} Gros, A., Goldwurm, A., Cadolle-Bel, M., et al. 2003, A\&A, 
	411, L179

\bibitem{} Halpern, J.P. 2005, ATel 572

\bibitem{} Halpern, J.P., \& Gotthelf, E.V. 2006, ATel 692

\bibitem{} Halpern, J.P., \& Mirabal, N. 2006, ATel 709

\bibitem{} Hamuy, M., Walker, A.R., Suntzeff, N.B., et al. 1992, 
	PASP, 104, 533

\bibitem{} Hamuy, M., Suntzeff, N.B., Heathcote, S.R., et al. 1994, 
	PASP, 106, 566

\bibitem{} Ho, L.C., Filippenko, A.V., \& Sargent, W.L.W. 1993, ApJ, 417,
        63

\bibitem{} Ho, L.C., Filippenko, A.V., \& Sargent, W.L.W. 1997, ApJS,
        112, 315

\bibitem{} Horne, K. 1986, PASP, 98, 609

\bibitem{} Kaspi, S., Smith, P.S., Netzer, H., et al. 2000, ApJ, 533, 631

\bibitem{} Kuiper, L., Keek, S., Hermsen, W., Jonker, P.G., \& Steeghs, D. 
	2006, ATel 684

\bibitem{} Lang, K.R. 1992, Astrophysical Data: Planets and Stars.
        Springer-Verlag, New York

\bibitem{} Leitherer, C. 1988, ApJ, 326, 356

\bibitem{} Masetti, N., Palazzi, E., Bassani, L., Malizia, A., \&
        Stephen, J.B. 2004, A\&A, 426, L41 (Paper I)

\bibitem{} Masetti, N., Mason, E., Bassani, L., et al. 2006a, A\&A, 
	448, 547 (Paper II)

\bibitem{} Masetti, N., Pretorius, M.L., Palazzi, E., et al. 2006b, A\&A, 
	449, 1139 (Paper III)

\bibitem{} Masetti, N., Bassani, L., Bazzano, A., et al. 2006c, A\&A, 
	455, 11 (Paper IV)

\bibitem{} Masetti, N., Morelli, L., Palazzi, E., et al. 2006d, ATel 783 

\bibitem{} Masetti, N., Bassani, L., Dean, A.J., Ubertini, P. \& Walter, 
	R. 2006e, ATel 715

\bibitem{} Masetti, N., Bassani, L., Bazzano, A., et al. 2006f, ATel 815

\bibitem{} Massey, P. 2002, ApJS, 141, 81

\bibitem{} Negueruela, I., \& Smith, D.M. 2006, ATel 831

\bibitem{} Osterbrock, D.E., \& Pogge, R.W. 1985, ApJ, 297, 166

\bibitem{} Osterbrock, D.E. 1989, Astrophysics of Gaseous Nebulae and
        Active Galactic Nuclei (Mill Valley: Univ. Science Books)

\bibitem{} Revnivtsev, M., Sazonov, S.Y., Molkov, S.V., et al. 2006a, 
	Astron. Lett., 32, 145

\bibitem{} Revnivtsev, M., Sazonov, S.Y., Churazov, E.M., \& Trudolyubov, 
	S. 2006b, A\&A, 448, L49

\bibitem{} {\it ROSAT} Team 2000, {\it ROSAT} News No. 71, The {\it ROSAT}
	Source Catalog of Pointed Observations with the High Resolution 
	Imager (1RXH; 3$^{\rm rd}$ Release)

\bibitem{} Sazonov, S.Y., Churazov, E., Revnivtsev, M.G., Vikhlinin, A.,
        \& Sunyaev, R.A. 2005, A\&A, 444, L37

\bibitem{} Schlegel, D.J., Finkbeiner, D.P., \& Davis, M. 1998, ApJ, 500,
        525

\bibitem{} Skrutskie, M.F., Cutri, R.M., Stiening, R., et al. 2006, AJ, 
	131, 1163

\bibitem{} Stephen, J.B., Bassani, L., Molina, M., et al. 2005, A\&A, 432, 
	L49

\bibitem{} Stephen, J.B., Bassani, L., Malizia, A., et al. 2006, A\&A, 
	445, 869

\bibitem{} Suleimanov, V., Revnivtsev, M., \& Ritter, H. 2005, A\&A, 435,
	191

\bibitem{} Tomsick, J.A., Chaty, S., Rodriguez, J., et al. 2006, ApJ,
	647, 1309

\bibitem{} Ubertini, P., Lebrun, F., Di Cocco, G., et al. 2003, A\&A,
        411, L131

\bibitem{} Veilleux, S., \& Osterbrock, D.E. 1987, ApJS, 63, 295

\bibitem{} Voges, W., Aschenbach, B., Boller, T., et al. 1999, A\&A, 349, 
	389

\bibitem{} Voges, W., Aschenbach, B., Boller, T., et al. 2000, IAU Circ. 
	7432

\bibitem{} Walter, R., Zurita Heras, J., Bassani, L., et al. 2006, A\&A,
        453, 133

\bibitem{} Warner, B. 1995, Cataclysmic variable stars (Cambridge:
        Cambridge Univ. Press)

\bibitem{} Wegner, W. 1994, MNRAS, 270, 229

\bibitem{} Winkler, H. 1992, MNRAS, 257, 677

\bibitem{} Winkler, C., Courvoisier, T.J.-L., Di Cocco, G., et al. 2003,
        A\&A, 411, L1

\bibitem{} Wu, X.-B., Wang, R., Kong, M.Z., Liu, F.K., \& Han, J.L. 2004,
        A\&A, 424, 793

\end{thebibliography}
\end{document}